\documentclass[aps,prb,twocolumn,superscriptaddress]{revtex4-2}

\usepackage{amsmath}
\usepackage{float}
\usepackage{graphicx}	
\usepackage{siunitx}
\usepackage{amsfonts}  
\usepackage{multirow}
\usepackage{amssymb}
\usepackage{soul}
\usepackage[table]{xcolor}
\usepackage{comment}
\usepackage{array}
\usepackage{tikz}
\usetikzlibrary{tikzmark}

\usepackage[colorlinks=true]{hyperref}	

\newcommand\Tstrut{\rule{0pt}{2.6ex}}         
\newcommand\Bstrut{\rule[-0.9ex]{0pt}{0pt}}   

\definecolor{Red}{RGB}{215,39,41}

\definecolor{Blue}{RGB}{31,118,181}
\begin{document}

	\title{Optimizing proximitized magnetic topological insulator nanoribbons for Majorana bound states}
	
    \author{Edu\'ard Zsurka}
    \email[]{eduard.zsurka@uni.lu}
    \affiliation{Peter Grünberg Institute (PGI-9), Forschungszentrum Jülich, 52425 Jülich, Germany}
    \affiliation{JARA-Fundamentals of Future Information Technology, Jülich-Aachen Research Alliance, Forschungszentrum Jülich and RWTH Aachen University, 52425 Jülich, Germany}
    \affiliation{Department of Physics and Materials Science, University of Luxembourg, 1511 Luxembourg, Luxembourg}
    \author{Daniele Di Miceli}
    \affiliation{Institute for Cross-Disciplinary Physics and Complex Systems IFISC (CSIC-UIB), E-07122 Palma, Spain}
    \affiliation{Department of Physics and Materials Science, University of Luxembourg, 1511 Luxembourg, Luxembourg}
    \author{Julian Legendre}
    \affiliation{Department of Physics and Materials Science, University of Luxembourg, 1511 Luxembourg, Luxembourg}    
    \author{Llorenç~Serra}
    \affiliation{Institute for Cross-Disciplinary Physics and Complex Systems IFISC (CSIC-UIB), E-07122 Palma, Spain}
    \affiliation{Department of Physics, University of the Balearic Islands, E-07122 Palma, Spain}
    \author{Detlev Gr\"utzmacher}
    \affiliation{Peter Grünberg Institute (PGI-9), Forschungszentrum Jülich, 52425 Jülich, Germany}
    \affiliation{JARA-Fundamentals of Future Information Technology, Jülich-Aachen Research Alliance, Forschungszentrum Jülich and RWTH Aachen University, 52425 Jülich, Germany}    
    \author{Thomas L. Schmidt}
    \affiliation{Department of Physics and Materials Science, University of Luxembourg, 1511 Luxembourg, Luxembourg}
    \author{Kristof Moors}
    \thanks{Present Address: Imec, Kapeldreef 75, 3001 Leuven, Belgium}
    \email[]{kristof.moors@imec.be}
    \affiliation{Peter Grünberg Institute (PGI-9), Forschungszentrum Jülich, 52425 Jülich, Germany}
    \affiliation{JARA-Fundamentals of Future Information Technology, Jülich-Aachen Research Alliance, Forschungszentrum Jülich and RWTH Aachen University, 52425 Jülich, Germany}
	
    \date{\today}

\begin{abstract}
Heterostructures comprised of a magnetic topological insulator (MTI) placed in the proximity of an $s$-wave superconductor have emerged as a platform for the practical realization of Majorana bound states (MBSs). More specifically, it has been theoretically predicted that MBS can appear in proximitized MTI nanoribbons (PNRs) in the quantum anomalous Hall regime. As with all MBS platforms, disorder and device imperfections can be detrimental to the formation of robust and well-separated MBSs that are suitable for fusion and braiding experiments.
Here, we identify the optimal conditions for obtaining a topological superconducting gap that is \emph{robust} against disorder, with spatially separated \emph{stable} MBSs in PNRs, and introduce a figure of merit that encompasses these conditions.
Particular attention is given to the thin-film limit of magnetic topological insulators (MTIs), where the hybridization of the surface states cannot be neglected, and to the role of electron-hole asymmetry in the low-energy physics of the system.
Based on our numerical results, we find that (1) MTI thin films that are normal (rather than quantum spin Hall) insulators for zero magnetization are favorable, (2) strong electron-hole asymmetry causes the stability and robustness of MBS to be very different for chemical potentials above or below the Dirac point, and (3) the magnetization strength should preferably be comparable to the hybridization or confinement energy of the surface states, whichever is largest.
\end{abstract}

\maketitle

\section{Introduction}\label{sec:int}


Magnetic topological insulators (MTIs) are a promising class of materials for studying the quantum anomalous Hall (QAH) effect~\cite{Yu2010,Chang2013,Wang2013,Kou2013,Wang_2015,Yasuda2017,Tokura2019,Sun_2019, Deng2020,Yasuda2020, Wang2021, Qiu_2022} and topological superconductivity~\cite{Wang2015_chiral,Zeng2018, Chen2018,Atanov2024}. 
The Bi$_2$Se$_3$ family of materials, consisting of Bi$_2$Se$_3$, Bi$_2$Te$_3$ and Sb$_2$Te$_3$, are 3D topological insulators (TIs) that host surface states at the interface with vacuum, with massless Dirac dispersion and spin-momentum locking. MTIs can be obtained by inducing an intrinsic magnetization into such a TI, by doping with a suitable transition metal, such as Cr, Fe, or V \cite{Yu2010}. 

A MTI that is confined to a quasi-one-dimensional nanoribbon geometry and also proximitized by a conventional $s$-wave superconductor, e.g., by depositing a superconducting material on top of the nanoribbon~\cite{Schueffelgen2019,Atanov2024}, can effectively become a one-dimensional (1D) $p$-wave superconductor~\cite{schnyder2008,tewari2012}. Such proximitized magnetic topological insulator nanoribbons (PNRs) host topologically protected zero-energy states at their ends, known as Majorana bound states (MBSs)~\cite{Zeng2018, Chen2018}. MBSs have potential applications in quantum computing and information processing as carriers of quantum information that can be stored and manipulated in a robust manner due to their nonabelian exchange statistics~\cite{Kitaev2001, Freedman2003, Alicea2011}.

Recently, significant effort has been dedicated to inducing superconductivity in MTIs~\cite{Mandal2022,uday2023induced,Yi2024,mandal2024,Sato2024,Yuan2024,Huang_2024}, to the experimental realization of submicron MTI nanoribbons (NRs)~\cite{Qiu_2022, Zhou2023} and even PNRs based on Cr-doped (Bi,Sb)$_2$Te$_3$~\cite{Atanov2024}.
%
It is worth noting that MTIs have mainly been established as a solid experimental platform for realizing the QAH effect, with a quantized Hall conductivity~\cite{Rienks2019}. For this regime, a large magnetization, which yields a large QAH gap and, consequently, strongly edge-localized states, is typically sought. However, for inducing a (topological) superconducting gap in PNRs, there must be sufficient overlap between the edge states localized on opposing edges~\cite{Burke2024}. These conditions do not generally yield a robustly quantized QAH effect.

Here, we focus specifically on the optimal conditions for MBS formation in PNRs. As theoretical results have been conflicting on the size and topology of the hybridization gap of TI thin films~\cite{Forster2015,Forster2016, Zsurka2024}, we opt for using an effective thin-film model~\cite{Lu2010, Shan_2010, shen_book}, with material-specific parameters that directly capture the features of the surface-state Dirac cone by fitting to ARPES measurements on TI thin films~\cite{Zsurka2024}. We take an in-depth look at the low-energy physics of (P)NRs, and identify the physical properties that are crucial for the stability and robustness of MBS formation.
We introduce a figure of merit to optimize these crucial properties of PNRs and pinpoint the best conditions in terms of the material type and film thickness, magnetization strength and chemical potential.
Notably, we find that electron-hole asymmetry , which has been neglected in previous works~\cite{Zeng2018, Chen2018, Burke2024}, can significantly alter the low-energy spectrum near the Dirac point and, correspondingly, the optimal conditions for MBS formation.

Our work is structured as follows. In Sec.~\ref{sec:models} we present the theoretical framework used for modelling PNRs. In Sec.~\ref{sec:LES} we introduce an analytical description of the low-energy states which appear in NRs, and study the effect of electron-hole asymmetry. In Sec.~\ref{sec:MBS} the essential quantities for the formation of MBS are presented and studied analytically. In Sec.~\ref{sec:results} the numerical calculations for realistic material parameters are presented, and compared to the analytical results. In Sec.~\ref{sec:discussion} we discuss additional relevant effects for topological superconductivity and MBSs in PNRs and in Sec.~\ref{sec:conclusion} we summarize our findings in a conclusion. For clarity, we included a table of all abbreviation we use in Appendix~\ref{app:abb}.

\section{Models}\label{sec:models}

\subsection{Continuum models}

The Bi$_2$Se$_3$ family of TIs are materials that consist of five-atom layers arranged along the $z$-direction, known as quintuple layers (QL). The electronic band structure around the $\Gamma$ point can be described using a four-band effective model, where only the orbitals which are responsible for the band inversion are considered. The \emph{bulk} Hamiltonian can be written in the following form~\cite{Zhang2009,Liu2010B}:
\begin{align}
\begin{split} H_\text{bulk}(\boldsymbol{k}) &= \varepsilon(\boldsymbol{k}) + \mathcal{M}(\boldsymbol{k})\tau_z + M_z\sigma_z \label{eq:H_3D} \\ 
&\hphantom{=} + B_0 k_z \tau_y + A_0(k_y\sigma_x-k_x\sigma_y)\tau_x, \end{split} \\
\varepsilon(\boldsymbol{k}) &\equiv C_0 + C_1 k_z^2 + C_2( k_x^2+k_y^2), \\
\mathcal{M}(\boldsymbol{k}) &\equiv M_0+M_1 k_z^2+M_2( k_x^2+k_y^2), 
\end{align}
with $\sigma_i$, $\tau_i$ ($i = x,y,z$) the Pauli matrices for the spin and orbital degree of freedom, respectively, and $A_0$, $B_0$, $C_{0,1,2}$ and $M_{0,1,2}$ parameters that can be determined from \emph{ab initio} calculations, for example~\cite{Zhang2009,Liu2010B,nechaev2016,Zsurka2024}. The effect of the ferromagnetically ordered, homogeneously distributed magnetic dopants is captured by the time reversal symmetry-breaking term $M_z \sigma_z$ with magnetization $M_z$ pointing in the out-of-plane direction. The considered materials
have large insulating gaps in the bulk, and gapless surface states on the interface with vacuum, ensured by a nontrivial $\mathbb{Z}_2$ topological invariant~\cite{Fu2008}.

When confined to a thin-film geometry, the surface states localized on opposite surfaces can hybridize, leading to a gap opening in the Dirac cone, around the Dirac point. In this limit, the bulk Hamiltonian in Eq.~\eqref{eq:H_3D} cannot always be used to describe (M)TI slabs in the thin-film limit~\cite{Zsurka2024}. Alternatively, the dispersion of the surface states in MTI thin films (along the $x\textnormal{-}y$ plane, i.e., parallel to the QLs) can be captured by the following effective \emph{thin-film} Hamiltonian~\cite{Lu2010}:
\begin{equation}
    \begin{split}
    H_\text{tf}(k_x,k_y) &= -Dk_{\parallel}^2 +v_\text{F}(k_y\sigma_x-k_x\sigma_y)\tau_z \label{eq:H_2D} \\
    &\hphantom{=} +(m_0-m_1k_{\parallel}^2)\tau_x + M_z\sigma_z,
    \end{split}
\end{equation}
with $\sigma_i$ ($i=x,y,z$) the Pauli matrices for spin, $\tau_{x,z}$ the Pauli matrices acting on the top/bottom subspace, corresponding to the surface states localized on top and bottom of the thin film, $k_{\parallel}^2 \equiv k_x^2+k_y^2$, and with $\hbar$ absorbed into $v_\text{F}$. The asymmetry of the Dirac cone is determined by $D$, $v_\text{F}$ is the Fermi velocity and the terms $m_0$ and $m_1$ determine the hybridization strength of the surface states. When setting $M_z=0$ we obtain the Bernevig-Hughes-Zhang model~\cite{bhz}, describing a normal insulator (NI) phase when $m_0m_1<0$ and a quantum spin Hall insulator (QSHI) phase when $m_0m_1>0$~\cite{Shan_2010, Lu2010, shen_book}. When $M_z\neq0$ and $|M_z|>|m_0|$ the system is in a quantum anomalous Hall insulator (QAHI) phase. Here, we focus on the effects of thin-film hybridization and electron-hole asymmetry, which are intrinsic to the materials under consideration. The advantages and disadvantages of the model Hamiltonians in Eqs.~\eqref{eq:H_3D} and~\eqref{eq:H_2D} have been discussed in Ref.~\cite{Zsurka2024} for TI nanostructures without magnetic dopants.

The effect of a superconducting material on top of a thin film can be described using the Bogoliubov-de Gennes (BdG) formalism, which yields the following Hamiltonian
\begin{eqnarray}
    H_\text{BdG}(\boldsymbol{k}) &=& \begin{pmatrix}
     H(\boldsymbol{k}) - \mu & i\Delta\sigma_y \\
     -i\Delta\sigma_y & - H^*(-\boldsymbol{k})+\mu  \\
    \end{pmatrix},
    \label{eq:H_BdG} \\
    \mathcal{H} &=& \frac{1}{2} \sum_{\boldsymbol{k}} \Psi_{\boldsymbol{k}}^\dagger H_\text{BdG}(\boldsymbol{k})
    \Psi_{\boldsymbol{k}},
    \label{eq:H_total}
\end{eqnarray}
with chemical potential $\mu$ and Nambu spinor $\Psi_{\boldsymbol{k}}^\dagger=(\boldsymbol{c}_{\boldsymbol{k}}^\dagger,\,\boldsymbol{c}_{-\boldsymbol{k}}^T) $ and $\boldsymbol{c}_{\boldsymbol{k}}^\dagger = (c_{\boldsymbol{k}A\uparrow}^\dagger, c_{\boldsymbol{k}B\uparrow}^\dagger, c_{\boldsymbol{k}A\downarrow}^\dagger, c_{\boldsymbol{k}B\downarrow}^\dagger)$, where $c_{\boldsymbol{k}i}^\dagger$ ($c_{\boldsymbol{k}i}$) are creation (annihilation) operators that form a basis for $H(\boldsymbol{k})$ ($-H(\boldsymbol{-k})^*$) with degrees of freedom $i\in\{A\uparrow,B\uparrow,A\downarrow,B\downarrow\}$. For the bulk Hamiltonian of Eq.~\eqref{eq:H_3D}, $A,B=\uparrow,\downarrow$ and $H(\mathbf{k}) = H_\text{bulk}(\mathbf{k})$ while, for the thin-film Hamiltonian of Eq.~\eqref{eq:H_2D}, $A,B=\text{t},\text{b}$ and $H(\boldsymbol{k}) = H_\text{tf}(k_x, k_y)$. In our calculations, we do not explicitly include the superconductor on top of the NR and only consider the proximity-induced pairing term that appears at the MTI nanoribbon-superconductor interface~\cite{Sitthison2014}. In the case of the bulk model of Eq.~\eqref{eq:H_3D} specifically, $\Delta$ is considered to be nonzero only in the top QL~\cite{Heffels2023}, while for the thin-film model of Eq.~\eqref{eq:H_2D} we consider $\Delta=\Delta_0(1+\tau_z)/2$ to represent proximity-induced pairing via the top surface~\cite{Chen2018}. We consider a conventional momentum-independent, positive real $s$-wave pairing potential $\Delta_0$, as there are no phase differences in our system.

\subsection{Lattice discretization}

In our simulations we employ a discretized version of the BdG Hamiltonian of Eq.~\eqref{eq:H_total} on a cubic (for the bulk Hamiltonian) or square (for the thin-film Hamiltonian) lattice, obtained via the finite-difference method implemented in the KWANT package~\cite{Groth2014}, which accurately captures the physics and topological properties of the PNR when the lattice constant is appropriately chosen. With these tight-binding models, we can consider arbitrarily shaped nanostructures (e.g., nanoribbons) in two and three dimensions. Here, we consider in particular ribbons along $x$, with width $W$ along $y$, and thickness along $z$ (in case of the bulk model), determined by the number of QLs with $N\, \text{QLs} \approx N\, \text{nm}$. We also use the Adaptive python package~\cite{Nijholt2019} for an effective sampling of the parameter space.

The proximity-induced superconducting gap induced in the nanoribbon is either topologically trivial or nontrivial. The topology of the gap of the quasi-1D PNR can be determined by evaluating
the $\mathbb{Z}_2$ topological invariant~\cite{Kitaev2001, Heffels2023},
\begin{equation}
    \mathcal{M} = \text{sign}\left\{\prod_{k_x=0,\pi/a}\text{Pf}[H_{\text{PNR}}(k_x)]\right\},
    \label{eq:maj_num}
\end{equation}
with Pf short for the Pfaffian and $H_{\text{PNR}}$ the tight-binding Hamiltonian over the cross section of the PNR, and wave number $k_x$ along the direction of the ribbon. The Pfaffian is evaluated with the pfapack package~\cite{wimmer2012algorithm}.

\section{Low-energy states}\label{sec:LES}

Prior to analysing PNRs, we study the low-energy states (LESs) that appear in NRs, when there is no superconductor placed on top of the nanoribbon. The dispersion of the surface states in the thin-film Hamiltonian of Eq.~\eqref{eq:H_2D} is given by
\begin{equation}\label{eq:E_SSG_k}
    E_\text{s}^{\pm,\theta}(k_x,k_y)=-D k_\parallel^2 + \theta\sqrt{v_\text{F}^2 k_\parallel^2 + (m_0 \pm M_z - m_1 k_\parallel^2 )^2 },
\end{equation}
where $\pm$ distinguishes two types of solutions (see Appendix~\ref{app:edge}) and $\theta=\pm$ differentiates between the positive ($+$) and negative ($-$) energy branches of the dispersion. At $k_x=k_y=0$, the LESs according to Eq.~\eqref{eq:E_SSG_k} have energies
\begin{equation}\label{eq:E_SSG}
    \varepsilon_\text{s}^\theta = \theta||m_0|- |M_z||,
\end{equation}
which define the surface-state gap (SSG). In a ribbon-like geometry, the surface states are quantized and, additionally, edge-state solutions can appear within the SSG. To this end, we first study in Sec.~\ref{sec:LESsym} the LESs in the electron-hole symmetric case, with $D=0$, and accordingly analyse only the positive-energy states. In particular, we determine the two LESs with the lowest positive energy, which we denote as $\text{LES}_1$ and $\text{LES}_2$. Subsequently, by setting $D\neq0$, in Sec.~\ref{sec:LESasym} we investigate the effect of electron-hole asymmetry on the LES (not to be confused with particle-hole symmetry, which generally applies for the BdG Hamiltonian). In what follows, our results were determined for $M_z>0$, but the $M_z<0$ case can be treated analogously. Additionally, as the topological properties of MTI thin films are determined by the relative sign of $m_0$ and $m_1$, we choose to set $m_1>0$ and use the sign of $m_0$ to differentiate between the QSHI and NI phases, without the loss of generality.

\subsection{Symmetric case ($D=0$)}\label{sec:LESsym}

We consider a semi-infinite plane geometry along $x\textnormal{-}y$ with $y\geq0$, and determine the edge-state solutions $\psi_\text{e}^\pm\propto \text{exp}(-y/\lambda^\pm+i k_x x )$, where $\lambda^\pm$ is the decay length into the bulk of the (M)TI and we impose $\psi_\text{e}^\pm(y=0)=\psi_\text{e}^\pm(y\to\infty)=0$~\cite{Zhang2016}. There are two edge-state solutions when the system is in the QSHI phase (here $m_0>0$ and $|M_z|<m_0$), corresponding to superscript $\pm$, and their decay length is equal to
\begin{equation}
    \lambda^\pm(k_x)=\dfrac{2m_1}{ v_\text{F}-\sqrt{ v_\text{F}^2-4m_1(m_0\pm M_z)+4m_1^2k_x^2}}\label{eq:kappa_pm}.
\end{equation}
When $v_\text{F}^2\gg 4m_1(m_0\pm M_z)$, we can approximate the decay lengths as
\begin{equation}
    \lambda^\pm(k_x)\approx\dfrac{v_\text{F}}{m_0 \pm M_z-m_1k_x^2}\label{eq:kappa_app_u}.
\end{equation}
By considering the approximation above, it becomes clear that, away from the $\Gamma$ point, $\lambda^\pm$ increases slightly with the magnitude of $k_x$. Also note that $\lambda^-$ ($\lambda^+$) diverges around the phase transition point $M_z\approx m_0$ ($M_z\approx-m_0$), when $m_0>0$ ($m_0<0$). In this regime, we cannot use $\psi_\text{e}^-$ ($\psi_\text{e}^+$) to approximate the LESs of the nanoribbon. However, for $M_z$ away from these phase transition points, the energy of the edge-state solutions, up to first order in $k_x$, is given by:
\begin{equation}\label{eq:eps_edge}
    \varepsilon_{\text{e}}^\pm(k_x)=\mp v_\text{F} k_x.
\end{equation}
We define the Dirac point (DP) energy $E_\text{DP}$ as the energy of the edge state in the semi-infinite plane at $k_x=0$. In the electron-hole symmetric case, $E_\text{DP}=0$. As this point will be relevant later, we note here that the edge states also do not carry a net spin $z$ polarization for any $k_x$~\cite{Zhou_2008}.

When the system is in the NI phase (here $m_0<0$ and $|M_z|<|m_0|$), there are no edge-state solutions, which is also reflected in the decay lengths of Eq.~\eqref{eq:kappa_app_u} becoming negative in this case. However, when magnetization overcomes the surface state hybridization strength $|M_z|>|m_0|$, the system enters the QAHI phase. For $M_z>0$, only one edge-state solution survives with inverse decay length $\lambda^+$ away from the edge. Note that, close to the phase transition $\lambda^+$ strongly depends on the sign of $m_0$.

\begin{figure}[t]
    \centering
    \includegraphics[width=\linewidth]{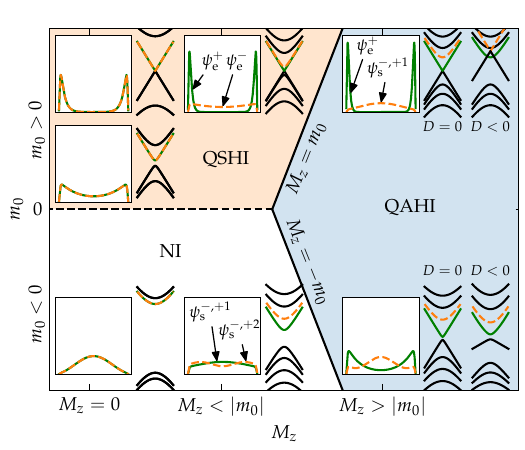}
    \caption{The wave function density $|\psi|^2$  at $k_x=0$, and the dispersion of the low-energy states $\text{LES}_1$ (green continuous line) and $\text{LES}_2$ (orange dashed line), for different values of the hybridization parameter $m_0$ and magnetization $M_z$, without electron-hole asymmetry $D=0$. In the insets, $|\psi|^2$ is shown as a function of the transverse coordinate $y$, and the label of the state that can be used to analytically describe the state. On the right side, we show the effect of the electron-hole asymmetry parameter $D$ on the dispersion of the NR. The different background colors represent the possible topological phases, with the NI ($0<M_z<-m_0$), QSHI ($0<M_z<m_0$) and QAHI ($|m_0|<M_z$) phases shown with a white, orange and blue background, respectively.}
    \label{fig:wf_map}
\end{figure}

We can approximate the edge states which appear at opposite sides of a MTI nanoribbon of width $W$, with the solutions obtained for the semi-infinite plane when the decay length of the edge states stays sufficiently below the ribbon width: $\lambda^\pm\ll W$. For example, for a 50 nm-wide nanoribbon, the edge-state solutions at $k_x=0$, $\psi_\text{e}^\pm$, satisfy $\lambda^\pm < W$ when $m_0\pm M_z>15$ meV. In Fig.~\ref{fig:wf_map}, we show the probability density for $k_x=0$ and the dispersion of the edge states and other LESs in a NR in qualitatively different regimes, when varying the parameters $m_0$ and $M_z$.
Note that we associate the term LES${}_{1(2)}$ with the entire first (second) lowest-energy subband of $k_x$-dependent solutions.

\begin{figure*}[t]
    \centering
    \includegraphics[width=\linewidth]{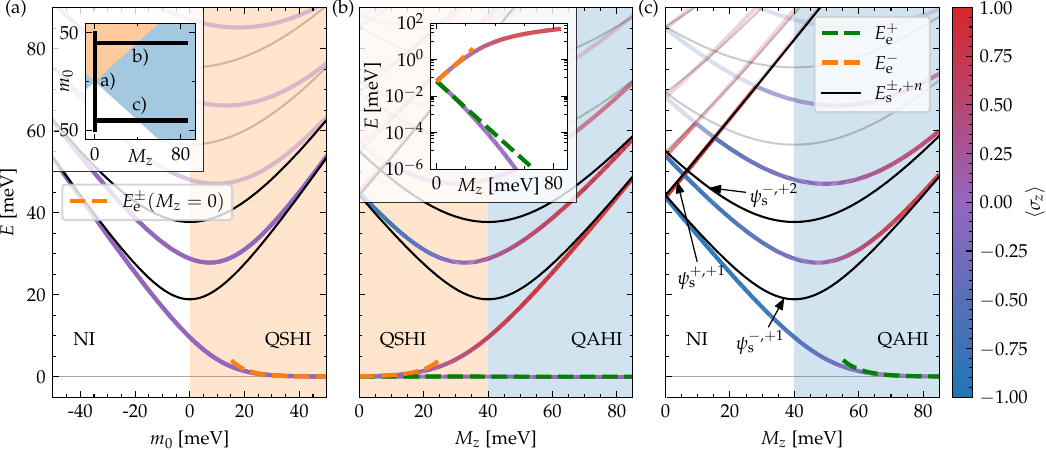}
    \caption{Low-energy states at $k_x=0$ in a 50 nm-wide MTI nanoribbon, as a function of the hybridization parameter $m_0$ or the magnetization $M_z$, with blue-to-red color scale indicating the (up-to-down) spin $z$ polarization of the states: (a) the nonmagnetic case ($M_z=0$) as a function of hybridization parameter $m_0$, with the sign of $m_0$ controlling whether the system is in the QSHI or NI phase; (b),(c) the magnetic case ($|m_0| = 40~\mathrm{meV}$) as a function of magnetization $M_z$, with (b) $m_0>0$, giving rise to a QSHI-to-QAHI crossover, and (c) $m_0<0$, giving rise to a NI-to-QAHI crossover.
    The dashed orange and yellow lines represent the approximate energy of the edge states (Eq.~\eqref{eq:E_edge}) and the continuous black lines represent the approximate energy of the confined surface states (Eq.~\eqref{eq:E_surf}).
    The inset in (a) shows the different parameter sweeps that are evaluated in $M_z\text{-}m_0$ space. The inset in (b) shows the two lowest energy states on a logarithmic scale in the QSHI phase. The background colors signify the topological phase, as already introduced in Fig.~\ref{fig:wf_map}. We consider $v_\mathrm{F}=3~\text{eV\AA}$, $m_1=15~\mathrm{eV\AA^2}$, $D=0~\mathrm{eV\AA^2}$.}
    \label{fig:lowest_energy}
\end{figure*}

In the finite-width geometry, the overlap of the edge states from opposite edges causes a small but relevant gap to appear (see Fig.~\ref{fig:wf_map}). This gap is due to edge-to-edge hybridization and not to be confused with the SSG introduced above. In the QSHI phase, the two edge states localized on one edge couple only to their counterpart on the opposite edge with the same decay length (see Fig.~\ref{fig:wf_map}). The edge states gap out at $k_x=0$ and their energy is shifted approximately from $0$ to $\pm E_{\text{e}}^\pm$, with
\begin{equation}\label{eq:E_edge}
    E_{\text{e}}^\pm \propto \, \text{exp}[-W/\lambda^\pm(k_x=0)].
\end{equation}
In Fig.~\ref{fig:lowest_energy}a we present the numerically determined spin-degenerate spectrum for a TI nanoribbon ($M_z=0$), and in Figs.~\ref{fig:lowest_energy}b,c the spectrum of a NR ($M_z\neq0$), for different signs of $m_0$. The analytical expression for the energy of the edge states according to Eq.~\eqref{eq:E_edge} shows good agreement with the numerical results.

In the QSHI phase, the low-energy states LES$_{1}$ and LES$_{2}$ correspond to the edge states $\psi_\text{e}^+$ and $\psi_\text{e}^-$, respectively, as we have $\lambda^+ < \lambda^-$ such that $E_\text{e}^+ < E_\text{e}^-$. 

\begin{table}[t]
    \centering
    \begin{ruledtabular}
    \begin{tabular}{c|ccccc}
    LES& QSHI & \multicolumn{2}{c}{NI} & \multicolumn{2}{c}{QAHI} \Bstrut\\
    \hline
    &\multicolumn{5}{c}{$D=0$} \Tstrut\Bstrut \\
    \hline
    $~~2$ &$\psi_\text{e}^-$ & \textcolor{Red} {$\psi_\text{s}^{+,+1}$}&\textcolor{Blue}{$\psi_\text{s}^{-,+2}$} & \multicolumn{2}{c}{\textcolor{Red}{$\psi_\text{s}^{-,+1}$}} \Tstrut\\
    $~~1$ &$\psi_\text{e}^+$   & \multicolumn{2}{c}{\textcolor{Blue}{$\psi_\text{s}^{-,+1}$}}   & \multicolumn{2}{c}{$\psi_\text{e}^+$}\Tstrut\\
    $-1$  &$\psi_\text{e}^+$   & \multicolumn{2}{c}{\textcolor{Red} {$\psi_\text{s}^{+,-1}$}}& \multicolumn{2}{c}{$\psi_\text{e}^+$}\Tstrut\\
    $-2$  &$\psi_\text{e}^-$   & \textcolor{Blue}{$\psi_\text{s}^{-,-1}$}&\textcolor{Red}{$\psi_\text{s}^{+,-2}$} & \multicolumn{2}{c}{\textcolor{Blue} {$\psi_\text{s}^{+,-1}$}}\Tstrut\Bstrut \\ [3pt]
    \hline
    &\multicolumn{5}{c}{$D\neq0$} \Tstrut\Bstrut \\
    \hline
    &&$M_z\lessapprox\frac{\eta^2}{|m_0|}$& $M_z\gtrapprox\frac{\eta^2}{|m_0|}$ & $m_0>0$ & $m_0<0$\Tstrut\Bstrut\\ [3pt]
    \hline
    $~~2$ &\textcolor{Red}{$\tilde{\psi}_\text{e}^+$}  & \textcolor{Red} {$\psi_\text{s}^{+,+1}$}&\textcolor{Blue}{$\psi_\text{s}^{-,+2}$}& \textcolor{Red}{$\tilde{\psi}_\text{e}^+$}&\textcolor{Red}{$\psi_\text{s}^{-,+1}$}\Tstrut \\
    $~~1$ &\textcolor{Blue}{$\tilde{\psi}_\text{e}^-$}& \multicolumn{2}{c}{\textcolor{Blue}{$\psi_\text{s}^{-,+1}$}}   &\textcolor{Red}{$\psi_\text{s}^{-,+1}$}& \textcolor{Red}{$\tilde{\psi}_\text{e}^+$}\Tstrut\\
    $-1$  &\textcolor{Red} {$\tilde{\psi}_\text{e}^+$}& \multicolumn{2}{c}{\textcolor{Red} {$\psi_\text{s}^{+,-1}$}} & \textcolor{Red}{$\tilde{\psi}_\text{e}^+$}& \textcolor{Red}{$\tilde{\psi}_\text{e}^+$}\Tstrut\\
    $-2  $&\textcolor{Blue}{$\tilde{\psi}_\text{e}^-$}  & \textcolor{Blue}{$\psi_\text{s}^{-,-1}$}&\textcolor{Red}{$\psi_\text{s}^{+,-2}$} & \textcolor{Blue} {$\psi_\text{s}^{+,-1}$}& \textcolor{Blue} {$\psi_\text{s}^{+,-1}$}\Tstrut\Bstrut \\
    \end{tabular}
    \end{ruledtabular}
    \caption{The order of the LESs and corresponding wave function in the electron-hole symmetric ($D=0$) and asymmetric case ($D<0$). The labels $2,1,-1$ and $-2$ indicate the ordering with respect to $E=0$ ($D=0$). In the asymmetric case, we use the same labels for the same states, albeit $E=0$ is no longer a good reference point. The colors show the net spin $z$ polarization of the state, with red (blue) corresponding to a net spin up (down) polarization.}
    \label{tab:LES}
\end{table}

In order to determine the LESs for the other phases, we have to consider higher-energy states that originate from the confinement of surface states to a ribbon geometry. These confined surface states can be treated as standing wave-like states in the transverse direction, with wave function $\psi_\text{s}^{\pm, n}\sim \text{sin}(n\pi y/W)\mathrm{exp}(i k_x x)$ and quantized transverse momenta $k_y \rightarrow n \pi /W$ ($n \in \mathbb{Z}_0$). We can use the dispersion of the surface states given in Eq.~\eqref{eq:E_SSG_k} and set $\theta=+1$ to approximate the energy of the confined surface states with $E>0$ at $k_x=0$ as~\cite{Bardarson2010}
\begin{equation}\label{eq:E_surf}
    E_\text{s}^{\pm,+n}\approx\sqrt{(|m_0|\pm M_z)^2+n^2\eta^2},
\end{equation}
where $\eta = v_\text{F}\pi/W$ is the confinement energy (e.g., for $v_\text{F}=3~\mathrm{eV\AA}$ and $W=50~\text{nm}$, we obtain $\eta\approx19~\text{meV}$). Note that we have ignored the terms quadratic in $k_y$ for obtaining this expression. At low magnetization, two states appear for each nonzero integer $n$, with opposite spin $z$ polarization -- one is pushed up, and the other is pushed down in energy by $M_z$. When $M_z>|m_0|$ however, all confined surface states at positive energies are spin up polarized along $z$ (see Fig.~\ref{fig:lowest_energy}). In general, a significant discrepancy between the analytical expression in Eq.~\eqref{eq:E_surf} and the numerical values for the confined surface states is observed only near the phase transition points (i.e., when $|m_0| \approx M_z$ ).

In the NI phase, no edge-state solutions are present (see Fig.~\ref{fig:wf_map}), thus the confined surface state $\psi_\text{s}^{-,+1}$ corresponds to $\text{LES}_1$. If the confinement energy dominates over $M_z$, more precisely, $M_z\lessapprox \eta^2/|m_0|$, $\psi_\text{s}^{+,+1}$ corresponds to $\text{LES}_2$ while, for higher magnetization, $\psi_\text{s}^{-,+2}$ becomes $\text{LES}_2$, as can be seen in Fig.~\ref{fig:lowest_energy}c.

In the QAHI phase, $\text{LES}_1$ is the edge-state solution $\psi_\text{e}^+$, which hybridizes with its counterpart on the opposite edge of the nanoribbon with energy $E_{\text{e}}^+$ (see Fig.~\ref{fig:wf_map}). Note, that the decay length $\tilde{\lambda}^+$ strongly depends on the sign of $m_0$ [see Eq.~\eqref{eq:E_edge}]. The $\text{LES}_2$ is a spin $z$ up-polarized surface state, with energy approaching $E_\text{s}^{-,+1}$. In Figs.~\ref{fig:lowest_energy}b and c we present how increasing $M_z$ drives the system from the QSHI and NI phase, respectively, into the QAHI phase.

In Table~\ref{tab:LES} we summarize the LESs for the different phases of the system. We also give the LESs for negative energies, where we label the first (second) LES closest to $E=0$ with $\text{LES}_{-1}$ ($\text{LES}_{-2}$), with energy opposite to that of $\text{LES}_1$ ($\text{LES}_2$).

\subsection{Asymmetric case  ($D\neq0$)}\label{sec:LESasym}

In most cases, the experimentally determined material parameters for the thin-film Hamiltonian of Eq.~\eqref{eq:H_2D} point to the presence of strong asymmetry ($|D|\approx|m_1|$), and also $D<0$~\cite{Zhang2010A, Zsurka2024}. A careful examination reveals that considering nonzero $D$ has a drastic effect on the LESs of NRs. Following the discussion of the previous section, we analyze how electron-hole asymmetry modifies the relevant features of the LESs, focusing on the first four LESs. 

First, the phase boundaries between the NI, QSHI and QAHI phases are not affected by asymmetry. Second, the decay length of the edge states $\tilde{\psi}^\pm$ in the QSHI phase (and $\tilde{\psi}^+$ in the QAHI phase) is modified to

\begin{equation}
    \begin{split}
        \tilde{\lambda}^\pm(k_x)&\approx\dfrac{v_\text{F}}{(m_0 \pm M_z-m_1k_x^2)\sqrt{1-\alpha^2}\mp \alpha v_\text{F} k_x},\\
    \end{split}
\end{equation}
where $\alpha=D/m_1$ and we also use a tilde above the symbols to denote the quantities derived in the asymmetric case. Note that this approximate expression is obtained in the limit $v_\text{F}^2\gg4m_1(m_0\pm M_z)(1-\alpha^2)$, which is satisfied for all the material parameters under consideration here, except 2 QL Sb$_2$Te$_3$ (see Appendix~\ref{app:params}). The energy of the edge states up to first order in $k_x$ is given by
\begin{equation}\label{eq:eps_edge_asym}
    \tilde{\varepsilon}_{\text{e}}^\pm(k_x)=-(m_0\pm M_z)\alpha\mp v_\text{F}\sqrt{1-\alpha^2} k_x.
\end{equation}
At $k_x=0$, the edge states no longer have zero energy, as the energy of the DPs is shifted to $E_\text{DP}^\pm=-(m_0\pm M_z)\alpha$~\cite{Legendre2024}. As $M_z$ increases, $E_\text{DP}^\pm$ become separated and $\tilde{\psi}_\text{e}^+$ ($\tilde{\psi}_\text{e}^-$) is shifted up (down) in energy, which has significant consequences that we discuss below. Throughout the paper, we refer to the DP with energy $E_\text{DP}^\pm$ as $\text{DP}^\pm$. Additionally, the Fermi velocity is rescaled by $\sqrt{1-\alpha^2}$, which can become significantly smaller than 1 for strong asymmetry. More details on this calculation are given in Appendix~\ref{app:edge}.

In Fig.~\ref{fig:kappas} we show the $k_x$ dependence of the decay length $\tilde{\lambda}^+$ in the QAHI phase, when $m_0<0$. When $D=0$ (see Fig.~\ref{fig:kappas}a), the decay length slightly increases away from the $\Gamma$ point independently of the sign of $k_x$, as already mentioned in the previous section. In the case of strong asymmetry ($|\alpha|\to1$), however, we observe a much stronger $k_x$ dependence (see Fig.~\ref{fig:kappas}b). The decay length diverges when the energy of the edge state reaches $-(m_0+M_z)/\alpha$, which, for $\alpha\to-1$ (the relevant regime for realistic material parameters, see Appendix~\ref{app:params}), lies just above the top of the SSG at energy $\varepsilon_{\text{s},+}=m_0+M_z$ [see Eq.~\eqref{eq:E_SSG}]. Incidentally, $E_\text{DP}^+=-(m_0+M_z)\alpha$ is also approaching the top of the SSG. As a consequence, the decay length $\tilde{\lambda}^+$ for energies near and above $\text{DP}^+$ is generally much larger than for energies below $\text{DP}^+$. In Table~\ref{tab:lambdas} we have listed the approximate value of $\tilde{\lambda}^+$ at relevant energies, which we also show in Fig.~\ref{fig:kappas}d, as a function of asymmetry strength $D/m_1$. In the case of $\alpha\to-1$ ($\nu=1+\alpha\to0$), the decay length $\tilde{\lambda}^+$ diverges at the top of the SSG and at $\text{DP}^+$, and vanishes at energies below $\text{DP}^+$. The resulting behavior is starkly different from the electron-hole symmetric case, where the decay length stays roughly constant throughout the SSG.
Note that, while we considered the impact of asymmetry on the decay length $\lambda^+$ in the QAHI phase with $m_0 < 0$ here, a similar analysis can be carried out for $\lambda^+$ in the QAHI phase when $m_0 > 0$, or for both decay lengths $\lambda^\pm$ in the QSHI phase, with qualitatively similar results.

The edge states also acquire a net spin polarization along $z$, given by
\begin{equation}
    \langle\tilde{\psi}^\pm|\sigma_z|\tilde{\psi}^\pm\rangle = \mp \alpha,
\end{equation}
which is independent of $k_x$. In the case of $\tilde{\psi}_\text{e}^+$, the edge state becomes completely spin up (down) polarized as $|\alpha|\to1$, when $D<0$ ($D>0$), as shown in Fig.~\ref{fig:kappas}c. Note, while previous works have determined the spin polarization for the edge states analytically, they typically consider $D=0$ and a chiral edge state in the QAHI phase with no net spin $z$ polarization~\cite{Zhang2016,uday2023induced}.

\begin{figure}[t]
    \centering
    \includegraphics[width=\linewidth]{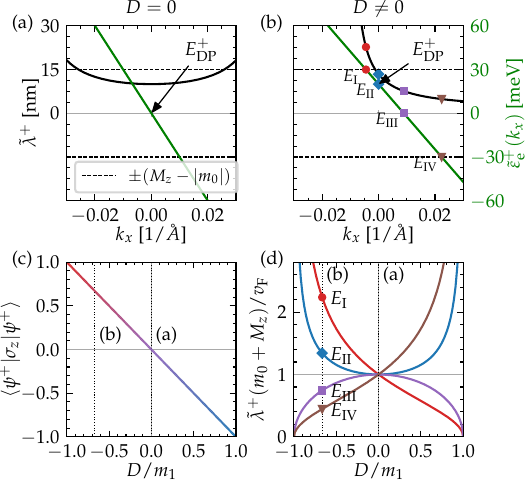}
    \caption{(a),(b) The energy $\varepsilon_\text{e}^+$ and decay length $\tilde{\lambda}^+$ of the edge state in the QAHI phase as a function of $k_x$, with (a) no electron-hole asymmetry ($D=0$) and (b) with electron-hole asymmetry ($D=-10~\mathrm{eV\AA^2}$). The gap edges are shown with dashed lines and the Dirac point energy $E_\text{DP}^+$ is indicated.
    (c) The spin $z$ polarization of the edge state as a function of the asymmetry parameter $D$. 
    (d) The decay length $\tilde{\lambda}^+$ as a function of asymmetry parameter $D$ for different energies, as listed in Table~\ref{tab:lambdas} and indicated in (b).
    We consider $v_\mathrm{F}=3~\text{eV\AA}$, $m_1=15~\mathrm{eV\AA^2}$, $m_0=-10~\mathrm{meV}$, and $M_z=40~\mathrm{meV}$.}
    \label{fig:kappas}
\end{figure}

\begin{table}[t]
    \centering
    \begin{ruledtabular}
    \begin{tabular}{c|cc}
    &$E$  & $\tilde{\lambda}^+(m_0+M_z)/v_\text{F}$\Bstrut\\ [5pt]
    \hline
    SSG top ($E_\text{I}$) & $M_z-|m_0|$ & $\sqrt{(2-\nu)/\nu}$\Tstrut\Bstrut\\ [3pt]
    $\text{DP}^+$ ($E_\text{II}$) & $E_{\text{DP}}^+$ & $1/(\sqrt{\nu(2-\nu)})$\Tstrut\Bstrut\\ [3pt]
    SSG center ($E_\text{III}$) &$0$ & $\sqrt{\nu(2-\nu)}$\Tstrut\Bstrut\\ [3pt]
    SSG bottom ($E_\text{IV}$) &$-M_z+|m_0|$ & $\sqrt{\nu/(2-\nu)}$\Tstrut\Bstrut\\ [3pt]
    \end{tabular}
    \end{ruledtabular}
    \caption{The decay length $\tilde{\lambda}^+$ at relevant energies and corresponding $k_x$ values, in the QAHI phase ($M_z>|m_0|$), when $m_0<0$, with $\nu \equiv 1+\alpha\equiv1+D/m_1$. The decay length $\tilde{\lambda}^-$ in the QSHI phase ($M_z<|m_0|$ and $m_0>0$) can be obtained by substituting $M_z\to-M_z$. In the case of $m_0>0$, $\tilde{\lambda}^+$ has a similar $m_0$ and $M_z$ dependence.}
    \label{tab:lambdas}
\end{table}

\begin{figure*}
    \centering
    \includegraphics[width=\linewidth]{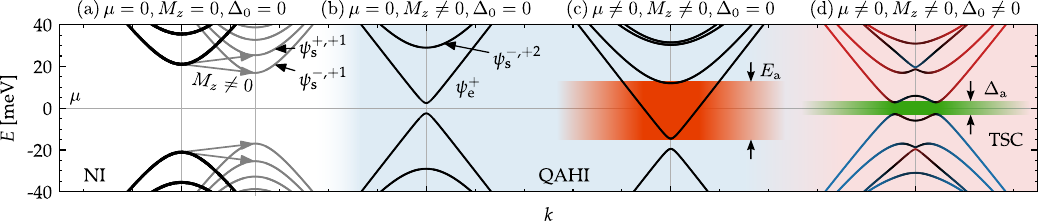}
    \caption{Schematic showing the realization of a PNR with MBS in steps, starting from a TI nanoribbon.
    (a) The dispersion of a TI nanoribbon in the NI phase has spin-degenerate bands, which in the presence of nonzero magnetization $M_z\neq0$ are split up (grey lines).
    (b) When $M_z>|m_0|$ the nanoribbon enters the QAHI phase, with the lowest energy state being an edge state.
    (c) The chemical potential $\mu$ is tuned to the center of the topological regime. The extent of the topological regime ($E_\text{a}$), indicated with orange shading, is an important quantity for the robustness of topological superconductivity throughout the PNR.
    (d) When proximitized by an $s$-wave superconductor, a superconducting gap ($\Delta_\text{a}$) opens up around zero energy. The size of this gap will determine the stability of the MBSs. The BdG spectrum is shown, with red (blue) lines corresponding to electron (hole)-like states, and the induced superconducting gap indicated with green shading. }
    \label{fig:pmtinr_components}
\end{figure*}

When considering a ribbon with finite width, the edge states localized at opposite edges hybridize, and cause an exponentially small gap to appear. In the QSHI phase, the hybridization gap opens up around $E_\text{DP}^\pm$ with energy $\tilde{E}_\text{e}^\pm$ for the edge states $\tilde{\psi}^\pm$ at $k_x=0$ that is proportional to
\begin{equation}\label{eq:E_edge_asym}
    \tilde{E}_{\text{e}}^\pm \propto\text{exp}[-W/\tilde{\lambda}^\pm(k_x=0)],
\end{equation}
considering $\lambda^\pm\ll W$. This results in the following energy for the $k = 0$ edge states,
\begin{equation}
    \tilde{E}_\text{e}^{\pm,\theta} = E_\text{DP}^\pm+\theta \tilde{E}_{\text{e}}^\pm,
\end{equation}
where, due to the absence of electron-hole symmetry, we use $\theta=\pm$ to differentiate between states that are shifted above ($\theta=+$) and below ($\theta=-$) the DP energy. In Table~\ref{tab:LES} we show the order of the LESs in ascending energy. In the previous section, we have labeled the LESs with positive (negative) index for positive (negative) energy. Here, with electron-hole asymmetry, $E=0$ is no longer a good reference point and we define $\text{LES}_{1,2}$ ( $\text{LES}_{-1,-2}$ ) more carefully by identifying the first two lowest (highest) energy states at $k_x=0$, with positive (negative) second derivative, i.e. $d^2 E(\text{LES}_{1,2})/dk_x^2>0$ $(<0)$. Note that, when $m_0>0$ (see first and second to last column), the order of the LESs changes to an alternating one (see Fig.~\ref{fig:wf_map}, when $D<0$). This effect can be attributed to the two pairs of edge states following the energies of their DP, which as mentioned previously, diverge as magnetization increases.

At higher energies, the confined surface states $\psi_\text{s}^{\pm,n}$ are present, with energy $E_\text{s}^{\pm,n}$ not significantly affected by electron-hole asymmetry, as described by Eq.~\eqref{eq:E_surf}. We also have to take into account the confined surface states originating from setting $\theta=-1$ in Eq.~\eqref{eq:E_SSG_k}, with energy
\begin{equation}
    E_\text{s}^{\pm,-n} \approx -\sqrt{(|m_0|\mp M_z)^2+n^2\eta^2},
\end{equation}
where the sign of $M_z$ has changed such that the energy of the state with $+$ ($-$) increases (decreases) with $M_z$ (when $M_z<|m_0|$).

In the NI case, the LESs are the same confined surface states as in the symmetric case (see Table~\ref{tab:LES}). 

In the QAHI phase, as in the symmetric case, the edge state $\tilde{\psi}_\text{e}^+$ couples to its counterpart on the opposing side of the NR. The energy of the two states at $k_x=0$ gaps out to $\tilde{E}_\text{e}^{+,+}$ ($\tilde{E}_\text{e}^{+,-}$), corresponding to $\text{LES}_1$ ($\text{LES}_{-1}$). Also $\text{LES}_{2,-2}$ remain confined surface states, with energy at $k_x=0$ roughly matching $E_\text{s}^{-,+1}$ and $E_\text{s}^{+,-1}$, respectively. Note that, when $m_0>0$ and asymmetry ($D<0$) is strong, the $\text{DP}^+$ edge states at $k_x=0$ can have higher energy than the lowest confined surface states with $\theta=+$ such that $\text{LES}_1$ ($\text{LES}_2$) now corresponds to a confined surface (edge) state, as shown in Table~\ref{tab:LES} (also see Appendix~\ref{app:edge}).

\begin{table*}
    \centering
    \begin{ruledtabular}
    \begin{tabular}{c|cc|cc|cc}
    & \multicolumn{2}{c|}{QSHI}& \multicolumn{2}{c|}{NI$^{(*)}$}  & \multicolumn{2}{c}{QAHI} \Bstrut\\ [3pt]
     \hline
    &(i) small $M_z$ &(ii) large $M_z$& (i) $M_z\lessapprox\eta ^2/|m_0|$& (ii) $M_z\gtrapprox\eta ^2\|m_0|$ & (i) $m_0>0$ & (ii) $m_0<0$ \Tstrut \Bstrut\\ [3pt]
    \hline
    $E_\text{a}$ &$\delta E_\text{DP}-\tilde{E}_{\text{e}}^-$ & 2$\tilde{E}_{\text{e}}^+$ & $2M_z$ & $3\eta^2/(2|m_0|)$&2$\tilde{E}_{\text{e}}^+$&$M_z-|m_0|-E_\text{DP}^+$\Tstrut \\
    $E_\text{b}$  &$\delta E_\text{DP}+\tilde{E}_{\text{e}}^-$& 2$\tilde{E}_{\text{e}}^-$ & $2M_z$ & $3\eta^2/(2|m_0|)$&$M_z-|m_0|+E_\text{DP}^+$&$M_z-|m_0|+E_\text{DP}^+$\Tstrut\Bstrut\\
    \hline
    \hline
    $\text{LES}_1~~$ &\multicolumn{2}{c|}{\textcolor{Blue}{$\tilde{\psi}_\text{e}^-$}}& \multicolumn{2}{c|}{\textcolor{Blue}{$\psi_\text{s}^{-,+1}$}} & \textcolor{Red}{$\psi_\text{s}^{-,+1}$}& \textcolor{Red}{$\tilde{\psi}_\text{e}^+$}\Tstrut\\
    $\text{LES}_{-1}$ &\multicolumn{2}{c|}{\textcolor{Red}{$\tilde{\psi}_\text{e}^+$}}& \multicolumn{2}{c|}{\textcolor{Red}{$\psi_\text{s}^{+,-1}$}}& \textcolor{Red}{$\tilde{\psi}_\text{e}^+$}& \textcolor{Red}{$\tilde{\psi}_\text{e}^+$}  \Tstrut\Bstrut\\
    \hline
    $\tilde{\lambda}_\text{a}$ & \multicolumn{2}{c|}{$\dfrac{v_\text{F}}{m_0-M_z}\dfrac{1}{\sqrt{\nu}}$} & \multicolumn{2}{c|}{-} & - & $\dfrac{v_\text{F}}{m_0+M_z}\dfrac{1}{\sqrt{\nu}}$  \Tstrut\\
    $\tilde{\lambda}_\text{b}$ & \multicolumn{2}{c|}{$\dfrac{v_\text{F}}{m_0}\dfrac{1}{\sqrt{2-\nu}}$$^{(**)}$} & \multicolumn{2}{c|}{-} & $\dfrac{v_\text{F}}{M_z}\dfrac{1}{\sqrt{2-\nu}}$$^{(***)}$ & $\dfrac{v_\text{F}}{m_0+M_z}\dfrac{1}{\sqrt{2-\nu}}$  \Tstrut\\
    \end{tabular}
    \end{ruledtabular}
    \caption{Analytical estimations of the size of the topological regimes (TRs) $E_\text{a,b}$ [see Eqs.~\eqref{eq:TR1},\eqref{eq:TR2}] and the decay length $\lambda_\text{a,b}$ of the corresponding edge states with chemical potential in the center of the TR [see Eqs.~\eqref{eq:lambda_a},\eqref{eq:lambda_b}], considering strong asymmetry ($|D|\approx|m_1|$) with $D<0$, which is equivalent to $\nu=1+D/m_1\to0$. The color of $\text{LES}_{1,-1}$ shows the net spin $z$ polarization of the state, with red (blue) corresponding to a net spin up (down) polarization. The edge-state solution $\psi_\text{e}^\pm$ is a valid approximation only when the condition $m_0\pm M_z\gg\eta$ is satisfied (see Sec.~\ref{sec:LES}). The transition between QSHI$_\text{i}$ and QSHI$_\text{ii}$ is given by solving $E_\text{DP}^+-E_\text{e}^+=E_\text{DP}^-+E_\text{e}^-$ for $M_z$.
    $^{(*)}$When $|m_0|\gg\eta$.
    $^{(**)}$When $m_0\gg M_z$.
    $^{(***)}$When $m_0\ll M_z$.}
    \label{tab:analytical}
\end{table*}

\section{Majorana bound states}\label{sec:MBS}

In this section, we switch to the PNR heterostructure, where an $s$-wave superconductor is placed on top of the NR. We identify in Sec.~\ref{sec:MBS_rs} the two essential quantities in PNRs for hosting robust and stable MBS. Building on our findings regarding the properties of the LESs, in Sec.~\ref{sec:MBS_app} we determine approximate analytical expressions that provide insight into the optimal conditions for MBSs.

\subsection{Robustness and stability}\label{sec:MBS_rs}
In a PNR heterostructure, superconducting correlations are induced into the states of the MTI with energy close to the chemical potential $\mu$ and give rise to topological superconductivity (note that Fermi level and chemical potential can be used interchangeably in this context). In particular, we analyze the crucial properties of PNRs for the formation of MBSs. To this end we present schematically in Fig.~\ref{fig:pmtinr_components} the dispersion of a 3D TI nanoribbon and we analyze how it changes when all the necessary components for the formation of MBSs are included. 

First, in Fig.~\ref{fig:pmtinr_components}a, we consider the doubly-degenerate dispersion a TI nanoribbon in the NI phase, with a sizeable hybridization gap. The energy of the confined surface states can be approximated by $E_\text{s}^{\pm,n}$ [see Eq.~\eqref{eq:E_surf}]. The addition of magnetic dopants lifts the time-reversal symmetry and the corresponding twofold degeneracy of the subbands, as shown for $|M_z|<m_0$ by the grey subbands in Fig.~\ref{fig:pmtinr_components}a. When magnetization overcomes the hybridization strength, the system enters the QAHI phase and within the SSG the edge state $\psi_\text{e}^+$ appears, as shown in Fig.~\ref{fig:pmtinr_components}b. 

At this point, we note that the number of MBSs appearing at the ends of a PNR is equal to the number of subbands of the non-superconducting NR crossing $\mu$ (at $E=0$). However, the PNR has a nontrivial topological invariant with a single stable MBS at each end of the ribbon only if an odd number of subbands are crossing zero energy, while pairs of MBS appearing at the same end are not protected against disorder~\cite{Potter2010, Stanescu2011, tewari2012, Heffels2023, Burke2024}. In this work, we focus on the case when a single subband crosses $E=0$ and a single MBS appears at each end of the PNR. Consequently, we identify the energy range where only a single subband crosses $E=0$, which we refer to as the topological regime (TR). In Fig.~\ref{fig:pmtinr_components}c the chemical potential $\mu$ is shifted with respect to Fig.~\ref{fig:pmtinr_components}b such that $E=0$ is intersected by the edge state $\psi_\text{e}^+$. The size of the TR in this case is $E_\text{a}=E_\text{s}^{-,+1} -E_\text{e}^+$.

Additionally, due to the dispersion of the NR, the number of subbands crossing $E=0$ increases as the chemical potential is further away above or below $\text{DP}^+$ (see Fig.~\ref{fig:pmtinr_components}). Thus, we distinguish two TRs, the one above $\text{DP}^+$ with range $E_\text{a}$, and the one below $\text{DP}^+$ with range $E_\text{b}$. Note that the TR above $\text{DP}^+$ rather appears around $\text{DP}^+$ in one specific case, namely in the QAHI phase with $m_0 < 0$ and strong asymmetry (i.e., the case with $\text{LES}_1$ being a confined surface state, see top right of Fig.~\ref{fig:wf_map} and Appendix~\ref{app:edge}). For convenience, we generally refer to the TR with range $E_\text{a}$ as the TR above $\text{DP}^+$ below.

In real PNR systems, there will be some amount of disorder, which can be understood effectively as local chemical potential variations $ \mu \to \mu + \delta\mu(\boldsymbol{r})$. Such fluctuations can shift the chemical potential to such a degree, that the number of subbands crossing $E=0$ changes along the length of the PNR, and additional MBSs appear, which then hybridize with the end-localized MBSs. In order to ensure the robustness of the topological regime with end-localized MBSs against such detrimental effects, we identify the \emph{size of the TR} as the first property of PNRs that should be maximized.

Next, we consider the proximity-induced superconducting gap (PG). In Fig.~\ref{fig:pmtinr_components}d we show the BdG spectrum of a PNR, with red (blue) representing electron (hole)-like subbands. A proximity-induced topological gap $\Delta_\text{ind}$ opens up when the chemical potential $\mu$ is tuned within a sufficiently large TR. The localization length of the MBS $\xi \approx v_\text{F}/ \Delta_\text{ind}$ is inversely proportional to $\Delta_\text{ind}$. A larger PG, and thus a shorter localization length, is beneficial for the stability of MBSs, as it ensures a good separation of MBSs at opposite ends of the PNR ($L \gg \xi$) and offers better protection against quasiparticle poisoning~\cite{Rainis2012}. Consequently, we consider the \emph{size of the PG} as the second property of PNRs that should be maximized.

We have identified the two crucial properties, the size of the TR to ensure the robustness of topological superconductivity, and a large PG for the stability of MBSs in PNRs. Next, we want to simultaneously maximize these properties for any given set of parameters of the PNR. To this end, we introduce a dimensionless figure of merit (FOM) that we can maximize, defined as
\begin{equation}
\text{FOM}_\beta \equiv \sqrt{\dfrac{\Delta_\beta}{\Delta_0}\dfrac{ E_\beta}{100~\text{meV}} }.
\label{eq:fom}
\end{equation}
The FOM comes in two flavors, with $\beta=\text{a,\,(b)}$ denoting the TR above (below) $\text{DP}^+$, and $\Delta_\beta$ the size of the PG, evaluated for chemical potential $\mu$ tuned to the center of the corresponding TR.

\subsection{Analytical approximations}\label{sec:MBS_app}

Using what we learned in Sec.~\ref{sec:LES} about the physics of the LESs, we can estimate the size of the TR and strength of the PG with analytical expressions. In Table~\ref{tab:analytical}, we summarize the size of the TRs and decay length of the edge states in different regimes where they can be determined analytically, for strong electron-hole asymmetry $|D|\approx|m_1|$ and $D<0$. We refer to the different ranges of the parameters of the system that we consider as $\text{PHASE}_{\text{case}}$, with $\text{PHASE}=\text{QSHI,\,NI}$ or $\text{QAHI}$, and $\text{case}=\text{i}$ or $\text{ii}$.

The size of the TR above (below) $\text{DP}^+$ is determined by the difference between the energy of $\text{LES}_1$ ($\text{LES}_{-1}$) and the energy of $\text{LES}_2$ ($\text{LES}_{-2}$) at $k_x=0$, or
\begin{eqnarray}
    E_\text{a}=\;&E&(\text{LES}_2)|_{k_x=0}-E(\text{LES}_1)|_{k_x=0},\label{eq:TR1}\\
    E_\text{b}=\;&E&(\text{LES}_{-1})|_{k_x=0}-E(\text{LES}_{-2})|_{k_x=0}.\label{eq:TR2}
\end{eqnarray}
When $m_0>0$, in the QSHI or QAHI phase, the energy at $k_x=0$ of $\text{LES}_1$ can be lower than $\text{LES}_{-1}$ for strong enough magnetization. In the QSHI phase particularly, this occurs when $E_\text{DP}^+-E_\text{e}^+>E_\text{DP}^-+E_\text{e}^-$. The value of $M_z$ where this transition occurs is determined through the interplay of the asymmetry strength (controlled by $D$ and $m_1$), the hybridization parameter $m_0$ and the nanoribbon width $W$. In this situation, in order to correctly evaluate the size of the TRs, one has to switch $\text{LES}_1$ and $\text{LES}_{-1}$ in Eqs.~\eqref{eq:TR1} and~\eqref{eq:TR2}  (for more details see Appendix~\ref{app:qshi}).

The size of the TRs are determined by evaluating Eqs.~\eqref{eq:TR1} and~\eqref{eq:TR2}. From the contents of Table~\ref{tab:analytical} we can conclude that, in the NI phase, there should be no significant difference between the TRs above and below $\text{DP}^+$. However, in the QAHI and QSHI phases, we should expect $E_\text{a}\ll E_\text{b}$. 

The size of the PG will depend on whether the proximitized state is an edge state or a confined surface state. In case of the latter, the PG can be approximated by~\cite{Heffels2023}
\begin{equation}\label{eq:Dind_surf}
    \Delta_\text{ind} \approx \Delta_0/2.
\end{equation}
However, when the proximitized state is an edge state (with $\lambda \ll W$), we can consider
\begin{equation}\label{eq:Dind_edge}
    \Delta_\text{ind}\approx \frac{W}{\lambda}e^{-W/\lambda}\sqrt{1-\alpha^2}\Delta_0,
\end{equation}
where $\lambda$ is the decay length of the edge state crossing $E=0$. We determine the size of the PG in the different TRs by setting the chemical potential $\mu$ to the center of the TR. When $\mu$ lies in the center of $E_\text{a}$ ($E_\text{b}$) the state crossing $E=0$ corresponds to $\text{LES}_1$ ($\text{LES}_{-1}$) and the PG is $\Delta_\text{a}$ ($\Delta_\text{b}$). Consequently, in the QSHI or QAHI phases, the crucial parameter determining the size of the PG is the decay length of the edge state. To this end, we can approximate the decay length of the edge state in the TR above and below $\text{DP}^+$ as
\begin{eqnarray}
    \tilde{\lambda}_\text{a}&\approx&\sqrt{\tilde{\lambda}(E_\text{I})\tilde{\lambda}(E_\text{II})},\label{eq:lambda_a}\\ \tilde{\lambda}_\text{b}&\approx&\sqrt{\tilde{\lambda}(E_\text{II})\tilde{\lambda}(E_\text{IV})},\label{eq:lambda_b}
\end{eqnarray}
where $\tilde{\lambda}(E_\text{I}),\,\tilde{\lambda}(E_\text{II})$ and $\tilde{\lambda}(E_\text{IV})$ represent the decay length of the edge state at the SSG top, $\text{DP}^+$ and SSG bottom, respectively (see Fig.~\ref{fig:kappas} and Table~\ref{tab:lambdas}). Simply put, for the TR above [below] $\text{DP}^+$, we take the geometric average of the decay length at $\text{DP}^+$, $\tilde{\lambda}(E_\text{I.})$, and the decay length at the top [bottom] of the SSG, $\tilde{\lambda}(E_\text{II})$ [$\tilde{\lambda}(E_\text{IV})$].

Examining the analytical results, we can reach several important conclusions regarding the PG. When the state at $E=0$ is a confined surface state (for the NI phase and QAHI phase with $m_0>0$, above DP$^+$), the PG is generally expected to be large, as the states with $k_x$ and $-k_x$ have a good spatial overlap. In the other cases (QSHI phase and QAHI phase), however, the PG will depend on the decay length of the edge state [see Eq.~\eqref{eq:Dind_edge}], with approximate values given in Table~\ref{tab:analytical}. The edge state with momentum $k_x$ has to couple to the edge state with momentum $-k_x$, which is localized at the opposite edge of the nanoribbon~\cite{Burke2024}. Consequently, we expect a smaller PG when the state at $E=0$ is an edge state, than when it is a confined surface state. Furthermore, when asymmetry is strong ($|D|\approx|m_1|$) and $D<0$, we have $\lambda_\text{a} \gg \lambda_\text{b}$, and consequently $\Delta_\text{a}\gg \Delta_\text{b}$ in the QAHI and QSHI phases. In the QSHI phase, $\Delta_\text{a}$ is also expected to be large close to the phase transition ($M_z\approx|m_0|$), as the decay length diverges there.

Summarizing our analytical findings, we expect a good FOM in the NI phase, due to the state at $E=0$ being a surface state, independently of the position of $\mu$. In the QAHI and QSHI phases, however, we have simultaneously $E_\text{a}\ll E_\text{b}$ and $\Delta_\text{a}\gg \Delta_\text{b}$, which prevents us from drawing any clear conclusion on dependence of the FOM on the position of $\mu$. Nonetheless, as the state at $E=0$ is almost always an edge state in these cases, we generally expect the FOM to be lower as compared to the NI phase.

\section{Optimized conditions}\label{sec:results}

After having introduced our approach for optimizing PNRs for MBS and expectations based on analytical expressions, we present our numerical simulations results.
In subsection~\ref{sec:resultsA}, we demonstrate the consequences of electron-hole asymmetry, by extending the results of Ref.~\cite{Burke2024}. In this work, the thin-film model of Eq.~\eqref{eq:H_2D} was employed using model parameters without electron-hole asymmetry ($D=0$). We clarify the effect of asymmetry on the TR and PG.
In subsection~\ref{sec:resultsB}, we consider the material parameters derived for Bi$_2$Se$_3$, Bi$_2$Te$_3$ and Sb$_2$Te$_3$ thin films in Ref.~\cite{Zsurka2024} and determine the TR and the PG for a range of magnetization values and thicknesses. We evaluate the FOM and find the optimal parameter ranges for maximizing the robustness and stability of MBS. Lastly, in subsection~\ref{sec:resultsC} we test the FOM-based optimization by evaluating the low-energy spectrum of disordered PNRs with finite length.

\subsection{Impact of asymmetry}
\label{sec:resultsA}
\begin{figure}[t]
    \centering
    \includegraphics[width=\linewidth,trim={0 -.1cm 0 0}]{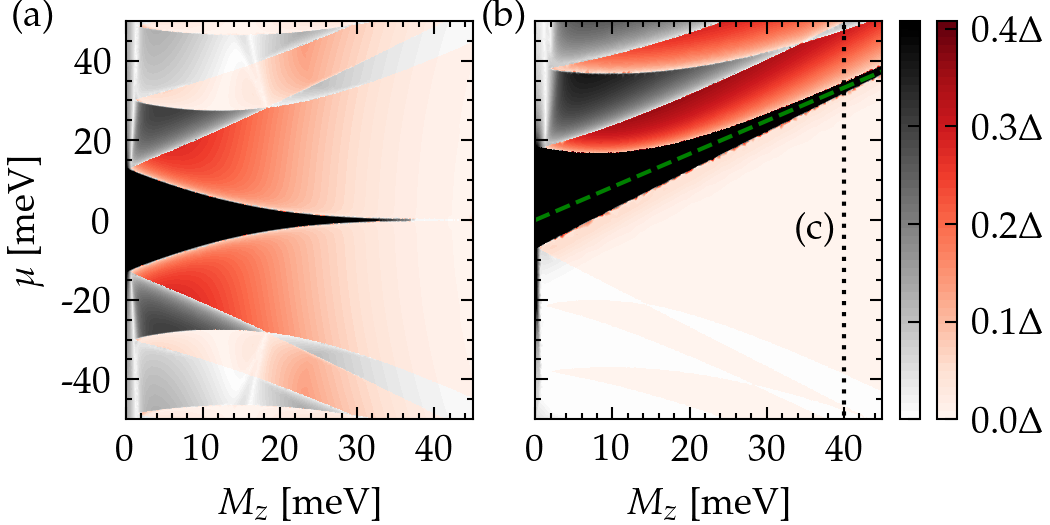}
    \includegraphics[width=\linewidth]{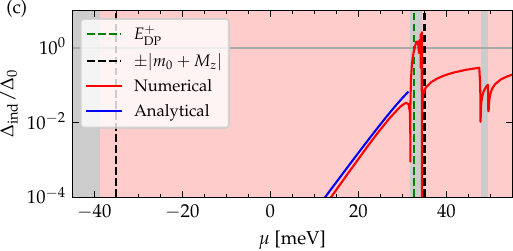}
    \caption{The size and topology of the proximity-induced superconducting gap $\Delta_\text{ind}$, determined numerically for a $W=50$ nm-wide PNR. The red (black) color map shows the size of the topological (trivial) gap in the case of (a) no asymmetry $D=0$ and (b) strong asymmetry $D=-14~\mathrm{eV\AA^2}$. In (c) we show a cross section of (b) taken at $M_z=40~\text{meV}$, and we compare the numerical result with the analytical formula in Eq.~\eqref{eq:Dind_edge}. We use $v_\mathrm{F}=3~\text{eV\AA}$, $m_1=15~\mathrm{eV\AA^2}$, $m_0=-5~\mathrm{meV}$~\cite{Burke2024}, and $\Delta_0=1~\mathrm{meV}$.}
    \label{fig:ind_diff}
\end{figure}

We emphasize the role of asymmetry in PNRs by evaluating $\Delta_\text{ind}$ in a 50 nm-wide nanoribbon, as in Fig.~1b of Ref.~\cite{Burke2024}. The size and topology of the PG, with trivial (nontrivial) regions indicated by a gray (red) color scale, is evaluated as a function of $\mu$ and $M_z$, for $D=0$ in Fig.~\ref{fig:ind_diff}a, and for $|D|\approx|m_1|$ and $D<0$ in Fig.~\ref{fig:ind_diff}b. 

When the system is electron-hole symmetric ($D=0$), the TR above and below $\text{DP}^+$, $E_\text{a}$ and $E_\text{b}$, represented by the two red colored wedges above and below $\mu=0$ in Fig.~\ref{fig:ind_diff}a, are identical for any value of the magnetization, $E_\text{a}(M_z)=E_\text{b}(M_z)$. The size of $\Delta_\text{ind}$ depends on the spatial distribution of the proximitized state. For $M_z<|m_0|$ ($m_0=-5~\text{meV}$) the nanoribbon is in the NI phase, with confined surface states that are delocalized over the width of the nanoribbon. For $M_z>|m_0|$ the nanoribbon enters the QAHI phase, and $\text{LES}_1$ and $\text{LES}_{-1}$ are edge states, which yield a weaker $\Delta_\text{ind}$ as $M_z$ is increased. The proximitized states also inherit the electron-hole symmetry of the Hamiltonian, and the size of the PG for chemical potential $\mu$ in the center of the TR is equal above and below $\text{DP}^+$, $\Delta_\text{a}(M_z) = \Delta_\text{b}(M_z)$.

In the presence of strong electron-hole asymmetry the TR and the PG are strongly affected, as shown in Fig.~\ref{fig:ind_diff}b. As outlined in Sec.~\ref{sec:MBS_app}, the TR above $\text{DP}^+$ gets reduced as $E_\text{DP}^+$ is pushed upwards with $M_z$, and the TR below $\text{DP}^+$ is enhanced. The size of the PG is also significantly affected by the presence of asymmetry. As explained in Sec.~\ref{sec:MBS_app}, the decay length of the edge states, and consequently the strength of $\Delta_\text{ind}$ is exponentially suppressed towards the SSG bottom and it reaches its maximal value close to the SSG top (see Fig.~\ref{fig:kappas}b). We take a line cut at $M_z=40$ meV, and compare the numerically extracted value of $\Delta_\text{ind}$ with the result of Eq.~\eqref{eq:Dind_edge} in Fig.~\ref{fig:ind_diff}c. The analytical result shows very good agreement with the numerically determined $\Delta_\text{ind}$.

\subsection{Material parameters}
\label{sec:resultsB}

In this section, we turn to analysing the FOM of PNRs with the material parameters for Bi$_2$Se$_3$, Bi$_2$Te$_3$ and Sb$_2$Te$_3$ in the thin-film limit, derived in Ref.~\cite{Zsurka2024}. The width of the nanoribbon is fixed to $W=50$ nm, and a realistic superconducting pairing of $\Delta_0=1$ meV is chosen~\cite{Odobesko2019}. We evaluate the relevant quantities for thickness $d=2-6$ QL, while varying the magnetization $M_z=0-70$ meV. We ignore 2 QL Bi$_2$Se$_3$ as the model parameters yield a nonphysical Fermi velocity.

\begin{figure*}
    \centering
    \includegraphics[width=\linewidth]{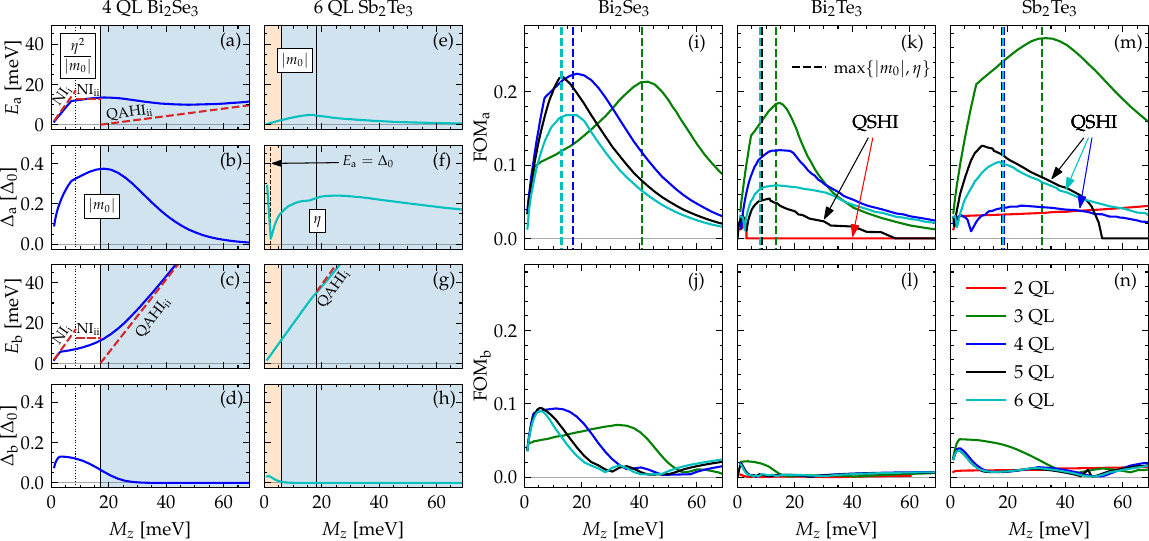}
    \caption{(a)-(h) The numerically determined topological regimes (TRs) $E_\mathrm{a,b}$ and induced superconducting gaps (PGs) $\Delta_\mathrm{a,b}$ for (a)-(d) 4 QL Bi$_2$Se$_3$ ($m_0<0$), for energies (a),(b) above ($E_\text{a}$, $\Delta_\text{a}$) and (c),(d) below ($E_\text{b}$, $\Delta_\text{b}$) the Dirac point ($\text{DP}^+$). Similarly for (e)-(h) 6 QL Sb$_2$Te$_3$ ($m_0>0$), the TR and PG (e),(f) above ($E_\text{a}$, $\Delta_\text{a}$) and (g),(h) below ($E_\text{b}$, $\Delta_\text{b}$) $\text{DP}^+$ are shown. Analytical approximations listed in Table~\ref{tab:analytical} are indicated for the TRs with red dashed lines. (i)-(n) The figure of merit of Eq.~\eqref{eq:fom} shown for the studied materials (i),(j) Bi$_2$Se$_3$, (k),(l) Bi$_2$Te$_3$, (m),(n) Sb$_2$Te$_3$ with the FOM determined (i),(k),(m) above $\text{DP}^+$ and (j),(l),(m) below the DP.}
    \label{fig:fom_full}
\end{figure*}

As expected from general considerations based on the analytical expressions in Sec.~\ref{sec:MBS_app}, we find that the sign of the hybridization parameter $m_0$ has a significant impact on the FOM. We analyze separately the TR and PG in two representative cases: in Figs.~\ref{fig:fom_full}a-d for 4 QL Bi$_2$Se$_2$, a NI (when $M_z=0$ and $\Delta_0=0$), and in Figs.~\ref{fig:fom_full}e-h for 6 QL Sb$_2$Te$_2$, a QSHI (when $M_z=0$ and $\Delta_0=0$).

In the case of $m_0<0$, we can approximate the size of the TR above and below $\text{DP}^+$ using Table~\ref{tab:analytical}. As shown in Figs.~\ref{fig:fom_full}a,c, for low values of magnetization $M_z\lessapprox\eta^2/|m_0|$ (NI$_\text{i}$), the TRs both increase linearly with the magnetization $E_\text{a,b}(M_z)\approx2M_z$. For higher values of the magnetization but still in the NI phase (NI$_\text{ii}$), the confinement energy $\eta$ determines the size of the TRs, with almost constant $E_\text{a,b}$ as a function of $M_z$. When $M_z$ overcomes the hybridization strength $m_0$, we can describe the size of the TRs using QAHI$_\text{ii}$, with $E_\text{a,b}$ again increasing linearly with $M_z$. In this regime we observe that $E_\text{a}\ll E_\text{b}$ as predicted in Sec.~\ref{sec:MBS_app}. The size of the PGs $\Delta_\text{a,b}$ is shown in Figs.~\ref{fig:fom_full}b,d. In the NI phase the state crossing $E=0$ is a surface state and in the QAHI phase it is an edge state (see Table~\ref{tab:analytical}), leading to $\Delta_\text{a,b}$ generally larger in the NI phase than in the QAHI phase. However, we also see that $\Delta_\text{a}$ is maximal around the phase transition between a NI and a QAHI. This effect can be explained by the fact that LES$_1$ changes its spin $z$ polarization from a down to an up polarized state (see Table~\ref{tab:analytical}), and around the phase transition point $M_z\approx|m_0|$ it has no net spin polarization, thus enhancing the PG. Note that, when $\eta > |m_0|$ (not the case here), LES$_1$ does not show a clear transition between a confined surface state and an edge state around $M_z\approx|m_0|$. In this case, we rather observe that FOM$_\text{a}$ peaks when the magnetization is comparable to the confinement energy $M_z\sim\eta$, where the decay length of the edge state is comparable to the width of the PNR, $\tilde{\lambda}\sim W$.

In the QAHI phase above $\text{DP}^+$, as we increase $M_z$ the edge state is increasingly edge localized leading to an exponentially suppressed $\Delta_\text{a}$ [see Eq.\eqref{eq:Dind_edge}] while, as a consequence of strong asymmetry, the decay length of the edge state is much shorter below than above $\text{DP}^+$, resulting in $\Delta_\text{a}(M_z>|m_0|)\gg \Delta_\text{b}(M_z>|m_0|)$.

Bringing together our observations for the TR and PG, we can determine the FOM for 4 QL Bi$_2$Se$_3$, as shown in Figs.~\ref{fig:fom_full}i,j. Above $\text{DP}^+$, FOM$_\text{a}$ peaks around the phase transition point $M_z\approx |m_0|$, where $\Delta_\text{a}$ also peaks, which can be understood considering that $E_\text{a}$ stays roughly constant throughout the range of $M_z$. Below $\text{DP}^+$ the FOM$_\text{b}$ is worse than above $\text{DP}^+$, due to the effect of electron-hole asymmetry on the decay length of the edge state.

Next, we turn to the $m_0>0$ case. The size of the TR in the QSHI above $\text{DP}^+$ can be described by the case QSHI$_\text{i}$. When $M_z>m_0$, we can mostly understand the size of the TR (Figs.~\ref{fig:fom_full}e,g) and the PG (Figs.~\ref{fig:fom_full}f,h) by considering case QAHI$_\text{ii}$ (see Table~\ref{tab:analytical}). The size of $E_\text{a}$ above $\text{DP}^+$ in Fig.~\ref{fig:fom_full}e is given by the exponentially small hybridization gap of the edge states while, below $\text{DP}^+$, in Fig.~\ref{fig:fom_full}g, $E_\text{b}$ similarly to the $m_0<0$ case increases linearly with $M_z$. The PG above $\text{DP}^+$ in Fig.~\ref{fig:fom_full}f is determined by the surface state crossing $E=0$, yielding an almost constant $\Delta_\text{a}$.
Interestingly, we observe that the PG above DP$^+$ peaks around the confinement energy, $M_z \sim \eta$ (similar to the $m_0<0$ case when $\eta > |m_0|$). We generally find that, for $M_z < \eta$, the PG increases with magnetization due to finite-size effects. When, however, $M_z$ is increased further above the confinement energy, the proximitized state LES${}_1$ becomes increasingly spin $z$ up polarized, leading to a decreasing PG. The interplay of these effects leads to a PG that is maximal in the intermediate regime when $M_z\sim\eta$.
Additionally, the PG below $\text{DP}^+$ is much smaller than the PG above $\text{DP}^+$, due to the state $\psi_\text{e}^+$ crossing $E=0$ being edge localized,
such that $\Delta_\text{a}(M_z>|m_0|) \gg \Delta_\text{b}(M_z>|m_0|)$.

We also determined the FOM for 6 QL Sb$_2$Te$_3$ in Figs.~\ref{fig:fom_full}k,l. Above $\text{DP}^+$, FOM$_\text{a}$ has a peak around the confinement energy $\eta$ coinciding with the peak in $\Delta_\text{a}$, and in comparison with the $m_0<0$ case (see Fig.~\ref{fig:fom_full}i), is much smaller due to the exponentially suppressed TR. Below $\text{DP}^+$ the FOM$_\text{b}$ is extremely small, reflecting the significantly smaller PG $\Delta_\text{a}\gg\Delta_\text{b}$.

Evaluating the FOM for all three considered materials in Fig.~\ref{fig:fom_full}i-n, we can draw a few conclusions regarding the optimal parameter values for MBSs in PNRs. We indeed see that (1) in the case of $m_0>0$ (2, 5 QL Bi$_2$Te$_3$ and 4-6 QL Sb$_2$Te$_3$) the FOM is comparatively worse than for $m_0<0$, even vanishing in some cases, due to higher-energy surface states (see Appendix~\ref{app:qshi} for details). Additionally, in the presence of strong asymmetry ($|D|\approx|m_1|$) with $D<0$ (being the case for all material parameters under consideration here, see~Appendix~\ref{app:params}) (2) FOM$_\text{a}$ above $\text{DP}^+$ (Figs~\ref{fig:fom_full}i-k) is always better than FOM$_\text{b}$ below $\text{DP}^+$ (Figs~\ref{fig:fom_full}l-n), as above $\text{DP}^+$ the proximitized state is either a surface state or an edge state with a much longer decay length than the edge state below $\text{DP}^+$. Finally, (3) the value of the FOM above DP$^+$ peaks close to the absolute value of the hybridization parameter $|m_0|$, where the state has no net spin $z$ polarization, when $|m_0|$ is larger than the confinement energy $\eta$. However, when $\eta>|m_0|$, we observe that the FOM is maximal around the confinement energy $\eta$ due to a combination of finite-size effects and properties of the LESs.

\subsection{Robustness and stability with disorder}
\label{sec:resultsC}

In experimentally realized MTIs, e.g. Cr-(Bi,Se)$_2$Te$_3$ samples, a Fermi level that aligns closely with the DP can be achieved by choosing the right ratio of Bi and Sb, which leads to inhomogeneously doped regions appearing throughout the device~\cite{Lee2015, chong2020, park2024}. These inhomogeneities (i.e., disorder) can effectively be described by local chemical potential fluctuations. In this section, we put our FOM to the test by studying numerically the spectra of disordered PNRs with finite length. We use a similar approach as described in Refs.~\cite{Heffels2023,Burke2024} to model chemical potential fluctuations $\delta \mu(\boldsymbol{r}) \equiv \overline{\delta \mu}\,\phi(\boldsymbol{r})$, with $\overline{\delta \mu}$ the fluctuation amplitude and $\phi(\boldsymbol{r})$ the spatial distribution given by a function with a Gaussian random-field profile
\begin{equation}\label{eq:disorder}
    \langle \phi(\boldsymbol{r}) \phi(\boldsymbol{r}') \rangle = \mathrm{exp\left[-(\boldsymbol{r}-\boldsymbol{r}')^2/(2\zeta^2)\right]},
\end{equation}
with correlation length $\zeta$. In our simulations we set to $\zeta=2$ nm, in agreement with experimentally resolved values~\cite{Beidenkopf_2011,Lee2015,chong2020}.

Here we also consider 50 nm-wide PNRs based on 4 QL Bi$_2$Se$_3$ and 6 QL Sb$_2$Te$_3$ (as in Figs.~\ref{fig:fom_full}a-h, with magnetization $M_z=20$ meV, within the range of experimentally determined values~\cite{Lee2015,chong2020,park2024}), which above the DP$^+$ have a relatively high and low $\text{FOM}_\text{a}$, respectively (see Figs.~\ref{fig:fom_full}i,m).

\begin{figure}[t]
    \centering
    \includegraphics[width=\linewidth]{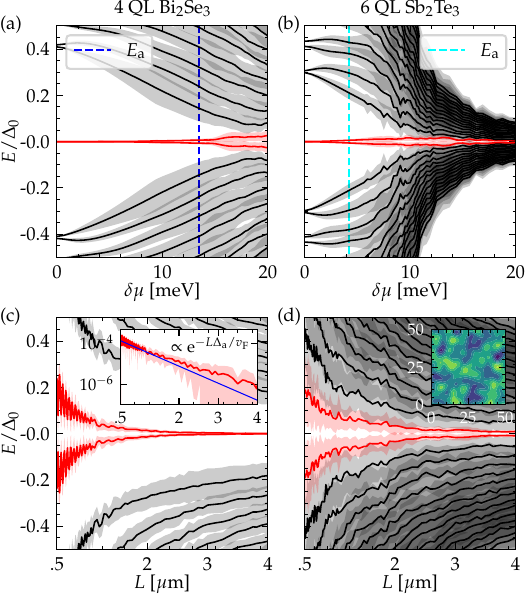}
    \caption{BdG spectra of 50 nm wide PNRs, constituting of (a),(c) 4 QL Bi$_2$Se$_3$ ($\text{FOM}_\text{a} \approx 0.22$) and (b),(d) 6 QL Sb$_2$Te$_3$ ($\text{FOM}_\text{a} \approx 0.1$), with magnetization $M_z=20$ meV. The spectra are determined as a function of disorder strength in (a), (b), for PNR length $L=3~\mathrm{\mu m}$, and as a function of PNR length in (c),(d), for fixed disorder strength $\overline{\delta \mu}=10$ meV. The size of the topological regime $E_\text{a}$ is shown with dashed lines. In the inset of (c) the energy of the MBS is shown on a logarithmic scale, with the analytical value overlaid. In the inset of (d) a 50x50 nm${}^2$ section of a disorder profile is shown, with correlation length $\zeta=2$ nm. Only the 30 lowest energy states are shown, averages were performed over 20 disorder profiles, and the standard deviation is indicated by shading.}
    \label{fig:disorder}
\end{figure}

To confirm the robustness of the topological gap with MBSs against disorder (determined by the size of the TR $E_\text{a}$), we vary the amplitude of the disorder $\overline{\delta \mu}$ and determine the low-energy spectrum in Figs.~\ref{fig:disorder}a,b. The length of the PNR is set to $L=3~\mathrm{\mu m}$, which is considerably larger than the localization length of the MBS, which for, e.g., 4 QL Bi$_2$Se$_3$ is $\xi\approx v_\text{F}/\Delta_\text{a}\approx 0.5~\mu\text{m}$. The two MBSs with energies around $E=0$ (shown with red lines), are localized at opposing ends of the PNR. In both cases, the MBSs start hybridizing when the disorder amplitudes approaches the size of the TR $\overline{\delta \mu}\approx E_\text{a}$. In the case of a low $\text{FOM}_\text{a}$, the topological gap vanishes and the MBSs hybridize for lower disorder amplitudes than in the case of a high $\text{FOM}_\text{a}$.

We also test the stability of MBSs (spatial separation determined by the size of the PG $\Delta_\text{a}$), by changing the length of the PNRs in the range $L=0.5-4~\mu \text{m}$, and show the spectrum of the PNRs in Figs.~\ref{fig:disorder}c,d. Here, the disorder amplitude is set to $\overline{\delta \mu}=10$ meV, which lies in the range of experimental observations~\cite{Beidenkopf_2011,nam2019,chong2020,park2024}. The hybridization gap of the MBS decreases as $\propto \mathrm{exp}(-L\Delta_\text{a}/v_\text{F})$ (see inset in Fig.~\ref{fig:disorder}c). When the TR is smaller than the disorder strength, however, the MBS pair couples to other low-energy states and is no longer topologically protected, as shown in Fig.~\ref{fig:disorder}d for a low $\text{FOM}_\text{a}$.

\section{Discussion}
\label{sec:discussion}

Throughout this work, the numerical calculations were performed for 50 nm-wide PNRs. Although ribbons with similar dimensions have been demonstrated, devices with wider MTI nanoribbons may be more easily fabricated. When considering wider nanoribbons, our findings (1), (2) and (3) still stand, however. We find that, due to the confinement energy being smaller, when $m_0<0$ (i.e. QSHI or QAHI phases) the surface states are closer in energy to the DP, and consequently the TR vanishes for lower magnetization values. This can also be seen for 5 QL Sb$_2$Te$_3$ in Fig.~\ref{fig:fom_full}m. The magnetization value $M_z$ where this happens decreases as the width of the PNR is increased. Thus, when considering wider PNRs, our conclusion (1), i.e., MTI nanoribbons with $m_0<0$ (i.e., ribbons in the NI phase when magnetization is absent) being better for the formation of MBS, becomes even more decisive.

In order to have a better understanding of our results, we also evaluated the FOM of thin-film Bi$_2$Se$_3$ and Bi$_2$Te$_3$ using the bulk Hamiltonian given in Eq.~\eqref{eq:H_3D}. Some of our results obtained with the thin-film Hamiltonian were reproduced qualitatively, such as (3) the FOM peak aligning with the hybridization or confinement energy. However, the (1) stark difference in FOM between NI and QSHI phases, or (2) between energies above and below the DP, was not reflected in the results obtained when using the bulk Hamiltonian. The observed discrepancies can be attributed to the fact that the material parameters used in the two models are obtained by fitting to different low-energy band structures. We discuss our findings in more detail in Appendix~\ref{app:3D}.

In experimental setups, PNRs are usually placed on a substrate or are interfaced with different materials, which can lead to interface effects (e.g., band bending) that are beyond the scope of our FOM analysis~\cite{menshov_2016_qah, menshov_2016_mod, wang2015, chong2023, chen2024}. Even in the setup that we have considered, the presence of a superconductor is expected to give rise to charge transfer across the interface with the MTI~\cite{ruessmann2022}. In this work, we have neglected such effects and focused on how the intrinsic PNR properties affect MBS formation. The insights from our FOM analysis remain relevant, however, even in the presence of such interface effects. Band bending that pushes the Fermi level away from the DP may be counteracted, for example, by electrostatic gating or adjusting the composition of the MTI close to the interface in such a way that band bending is compensated and the Fermi level remains close to the DP throughout the ribbon~\cite{lanius2016}.

\section{Conclusion}
\label{sec:conclusion}

In this work, we present a detailed analysis of PNRs, mainly focusing on the MTI thin films obtained by combining materials of the Bi$_2$Se$_3$ family, doped by transition metals, with a superconductor placed on top of the MTI, inducing superconducting correlations. We study this material platform with the intent of obtaining topological superconductivity, with a single pair of MBSs localized at opposing ends of the ribbon. To this end, we analyze the low-energy physics of PNRs and reveal the effect of electron-hole asymmetry on the system. We identify the relevant quantities for maximizing the stability and robustness of the topological superconducting regime and its MBSs. Based on analytical and numerical calculations, we pinpoint the optimal material properties and system parameters for the formation of MBSs in such heterostructures.

The optimal conditions for MBSs correspond to a large range of chemical potentials for the topological regime and a large proximity-induced superconducting gap. Our results indicate that, to maximize both these properties simultaneously, it is better to consider thin films of MTI materials (in particular magnetically doped thin films of the Bi$_2$Se$_3$ family of TIs) that (1) are NI ($m_0<0$) in the absence of magnetization and superconductivity, (2) the chemical potential should be near or above the Dirac point energy DP$^+$, and (3) the magnetization $M_z$ should be smaller or close to either the modulus of the hybridization parameter $|m_0|$ or the confinement energy of the surface states $\eta$, whichever is larger. The optimal PNR for hosting MBSs can be considered as a system with $m_0<0$ (see Table~\ref{tab:tf_params}), magnetization tuned close to the phase transition between the NI and QAHI phases, and the smallest possible width in order to maximize the topological regime with robust MBSs (see Table~\ref{tab:analytical}).

\begin{table}[b]
    \centering
    \begin{tabular}{c|c|cccc}
    \toprule
     &  $N_{\text{QL}}$ & $D~[\text{eV\AA}^2]$ & $m_0~[\text{eV}]$ & $m_1~[\text{eV\AA}^2]$ & $v_\text{F}~[\text{eV\AA}]$ \\
    \hline
    &     2 & -16.14 & -0.120 & 17.56 & -0.048 \\
    &     3 & -13.94 & -0.041 & 15.82 &  1.697 \\
    Bi$_2$Se$_3$
    &     4 & -13.31 & -0.017 & 16.35 &  1.920 \\
    &     5 & -13.23 & -0.007 & 16.51 &  2.010 \\
    &     6 & -13.06 & -0.003 & 16.75 &  2.046 \\
     \hline
    &     2 & -28.41 &  0.077 & 29.14 &  2.463 \\
    &     3 & -27.52 & -0.014 & 28.10 &  0.876 \\
    Bi$_2$Te$_3$
    &     4 & -29.55 & -0.002 & 30.06 &  1.301 \\
    &     5 & -28.50 &  0.003 & 28.97 &  1.340 \\
    &     6 & -28.36 &  0.000 & 28.81 &  1.232 \\
     \hline
    &     2 & -21.70 & -0.127 & 26.29 &  0.482 \\
    &     3 & -15.45 & -0.032 & 16.04 &  2.921 \\
    Sb$_2$Te$_3$
    &     4 & -13.39 &  0.004 & 13.89 &  2.952 \\
    &     5 & -12.65 &  0.009 & 13.10 &  2.870 \\
    &     6 & -13.38 &  0.006 & 13.83 &  2.887 \\
    \toprule
    \end{tabular}
    \caption{Parameters of the thin-film Hamiltonian [Eq.~\eqref{eq:H_2D}], for Bi$_2$Se$_3$, Bi$_2$Te$_3$ and Sb$_2$Te$_3$, for thicknesses ranging 2-6 QL, adapted from Ref.~\cite{Zsurka2024}.
}
    \label{tab:tf_params} 
\end{table}

\begin{acknowledgments}
	This work is supported by the QuantERA grant MAGMA, by the German Research Foundation under grant 491798118, and by MCIN/AEI/10.13039/501100011033 and the European Union NextGenerationEU/PRTR under project PCI2022-132927.
	K.M.\ acknowledges the financial support by the Bavarian Ministry of Economic Affairs, Regional Development and Energy within Bavaria’s High-Tech Agenda Project "Bausteine für das Quantencomputing auf Basis topologischer Materialien mit experimentellen und theoretischen Ansätzen" (Grant No.\ 07 02/686 58/1/21 1/22 2/23) and by the German Federal Ministry of Education and Research (BMBF) via the Quantum Future project ‘MajoranaChips’ (Grant No.\ 13N15264) within the funding program Photonic Research Germany.
\end{acknowledgments}

\appendix

\section{Material parameters}
\label{app:params}
The material parameters of the thin-film Hamiltonian [Eq.~\eqref{eq:H_2D}] were taken from Ref.~\cite{Zsurka2024} and are listed in Table~\ref{tab:tf_params}. The parameters describing the hybridization of the surface states are related as follows: $m_0=-\Delta/2$ and $m_1=-B$.

\section{Edge state in the presence of asymmetry}
\label{app:edge}
To understand the effect of the electron-hole asymmetry controlled by $D$ in the thin-film Hamiltonian of Eq.~\eqref{eq:H_2D}, we analyze the wave function of the system in a semi-infinite plane, similarly to the calculation outlined in Ref.~\cite{Zhang2016}. Starting from Eq.~\eqref{eq:H_2D}, we define a unitary transformation $U$,
\begin{equation}
U=\frac{1}{\sqrt{2}}\begin{pmatrix}
1 & 0 & 1 & 0\\0 & 1 & 0 & -1\\0 & 1 & 0 & 1\\1 & 0 & -1 & 0\nonumber
\end{pmatrix},
\end{equation}
which applied to $H_{\text{tf}}$ leads to
\begin{widetext}
\begin{equation}
H_\text{tf}'=UH_{\text{tf}}U^\dagger = 
\begin{pmatrix}
A_k + m_k + M_z  & i v_\text{F}k_- & 0 & 0\\
-i v_\text{F}k_+ & A_k - m_k - M_z & 0 & 0\\
0 & 0 & A_k + m_k - M_z & -i v_\text{F}k_+\\
0 & 0 & i v_\text{F}k_- & A_k - m_k + M_z
\end{pmatrix},
\label{eq:H_2D_2}
\end{equation}
\end{widetext}
where, for simplicity we defined $A_k \equiv -\mu-Dk^2$, $m_k \equiv m_0-m_1k^2$ and $k_\pm \equiv k_x\pm i k_y$. The topological properties of the system are determined by the signs of $m_0$, $m_1$ and $M_z$. In the case of the parameters chosen in Sec.~\ref{sec:LESasym} ($m_0<0$ and $M_z>|m_0|$), only the upper 2x2 block of $H'$ has an edge state solution, while the lower block has no solution.
We are interested in finding a solution for general energy $\varepsilon$ inside the band gap, for a system that semi-infinite in the $x-y$ plane, with an edge at $x=0$. We therefore take the ansatz
\begin{equation}\label{eq:ansatz}
    \tilde{\psi}_\text{e}^+(k_x,y)=\begin{pmatrix} c(k_x) \\ d(k_x) \end{pmatrix}\text{exp}(- \kappa^+ y),
\end{equation}
where $\kappa^+=1/\tilde{\lambda}^+$ is the inverse decay length. We consider the boundary conditions
\begin{align}
    \label{eq:boundary1}
    \tilde{\psi}_\text{e}^+(k_x,y\to0) = 0,\\
    \label{eq:boundary2}
    \tilde{\psi}_\text{e}^+(k_x,y\to\infty) = 0.
\end{align}
When $m_0<0$ and $M_z>|m_0|$ a band inversion occurs in the upper block of $H_\text{tf}'$ in Eq.~\eqref{eq:H_2D_2}. In the main text, the index $\pm$ differentiates between quantities determined when the band inversion happens in the upper ($+$) or lower ($-$) block of $H'$. Inserting the ansatz into $H_\text{tf}'$ we obtain
\begin{equation}\label{eq:H_upp}
\begin{split}
    H_{\text{upper}}(k_x,\kappa^+) &= -D(k_x^2-(\kappa^+)^2)+i v_\text{F} \sigma_x\kappa^+-v_\text{F}k_x\sigma_y \\
&+[m_0-m_1(k_x^2-(\kappa^+)^2)-M_z]\sigma_z.
\end{split}
\end{equation}
By solving the eigenvalue problem we can derive the relation between $\kappa^+$, $k_x$ and the energy $\tilde{\varepsilon}_\text{e}^+$ as
\begin{widetext}
\begin{equation}\label{eq:kappa_sq_asym_kx}
\begin{split}
    (\kappa^+)^2-k_x^2 &= \dfrac{v_\text{F}^2-2m_1(m_0+M_z)-2D\tilde{\varepsilon}_\text{e}^+}{2(m_1^2-D^2)}\\
    &\pm\dfrac{\sqrt{v_\text{F}^4 - 4v_\text{F}^2m_1(m_0+M_z)+4D^2(m_0+M_z)^2-4D\tilde{\varepsilon}_\text{e}^+(v_\text{F}^2-2m_1(m_0+M_z))+4m_1^2(\tilde{\varepsilon}_\text{e}^+)^2}}{2(m_1^2-D^2)}.
\end{split}
\end{equation}
\end{widetext}
The second b.c. in Eq.~\eqref{eq:boundary2} restricts $\kappa^+$ to positive values, which we label $\kappa^+_1$ and $\kappa^+_2$. Thus,  using the fact that both positive solutions $\kappa^+_1$ and $\kappa^+_2$ of Eq.~\eqref{eq:kappa_sq_asym_kx} are solutions of the eigenvalue problem we obtain the dispersion relation 
\begin{equation}\label{eq:E_u_asym}
\begin{split}
    \tilde{\varepsilon}_\text{e}^+ (k_x,\kappa^+_1,\kappa^+_2)&=(m_0+M_z)- (D+m_1)(\kappa^+_1 + \kappa^+_2) k_x \\ &-(D+m_1) \kappa^+_1 \kappa^+_2 - (D+m_1)k_x^2.
\end{split}
\end{equation}
As $\kappa^+_1$ and $\kappa^+_2$ are both $k_x$ dependent, we can determine the dispersion perturbatively up to first order $k_x$~\cite{shen_book} and we obtain Eq.~\eqref{eq:eps_edge_asym}. When the band inversion occurs in the lower 2x2 block of $H_\text{tf}'$ [see Eq.~\eqref{eq:H_2D_2}] for $m_0>0$ and $M_z<m_0$, one can similarly obtain the decay length $\lambda^-$ with dispersion given by $\tilde{\varepsilon}_\text{e}^-$.

\begin{figure}[t]
    \centering
    \includegraphics[width=\linewidth]{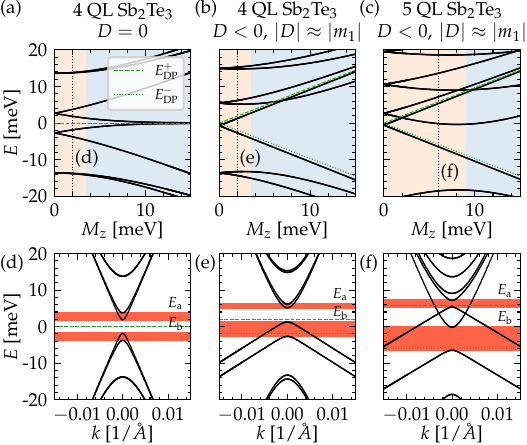}
    \caption{The dispersion of a 100 nm-wide nanoribbon for 4 QL Sb$_2$Te$_3$ (a),(d) without and (b),(e) with electron-hole asymmetry, and for (c),(f) 5 QL Sb$_2$Te$_3$. The dispersion is evaluated (a)-(c) at $k_x=0$ as a function of $M_z$ and (d)-(f) in the QSHI phase at selected magnetization values as a function of $k_x$. The energy of the DPs $E_\text{DP}^\pm$ are shown with green lines and in the bottom row we highlighted with orange the TR $E_\text{a}$ and $E_\text{b}$.}
    \label{fig:qshi}
\end{figure}

\begin{figure}[t]
    \centering
    \includegraphics[width=\linewidth]{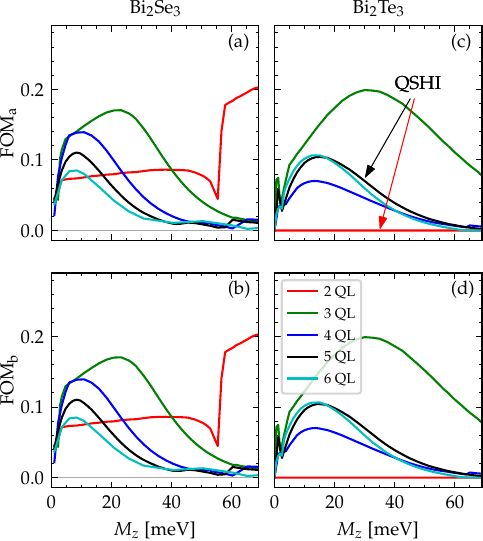}
    \caption{The figure of merit of Eq.~\eqref{eq:fom} shown for the studied materials (a),(b) Bi$_2$Se$_3$, (c),(d) Bi$_2$Te$_3$, with the FOM determined (a),(c) above the DP and (b),(d) below the DP. Here we use the bulk Hamiltonian [Eq.~\eqref{eq:H_3D}].}
    \label{fig:fom_full_3D}
\end{figure}

\section{Size of the topological regime when $m_0>0$}
\label{app:qshi}
As mentioned in the main text, when $m_0>0$ (QSHI phase with $M_z < m_0$) the size of the TRs may vary depending on the parameters of the system. The possible situations are presented in Fig.~\ref{fig:qshi}, with the dispersion shown in the electron-hole symmetric ($D=0$) case, in Figs.~\ref{fig:qshi}a,d, and in the strongly electron-hole asymmetric ($|D|\approx|m_1$, $D<0$) case, in Figs.~\ref{fig:qshi}b,c,e,f. In Sec.~\ref{sec:resultsB}, for all considered materials in the QSHI phase, the size of the TR can be described well by the QSHI$_\text{i}$ case (see Table~\ref{tab:analytical}), which can be observed in Fig.~\ref{fig:qshi}b. When the hybridization energy $\tilde{E}_\text{e}^-$ [see Eq.~\eqref{eq:E_edge_asym}] is small compared to the difference in energy of the DPs $\delta E_\text{DP}$, the TRs can be described well by the QSHI$_\text{ii}$ case, when, at $k_x=0$ the energy of LES$_1$ becomes lower than the energy of LES$_{-1}$, as presented in Fig.~\ref{fig:qshi}c,f. 

Additionally, in the QSHI or QAHI phases the energy of some surface states (such as $E_\text{s}^{-,+1}$) can become lower than the energy of the edge states $\tilde{E}_\text{e}^{+,\theta}$, which causes the FOM above $\text{DP}^+$ to vanish. This effect is captured in Fig.~\ref{fig:qshi}c at $M_z\approx m_0$, and is also the reason for $\text{FOM}_\text{a}$ vanishing in Figs.~\ref{fig:fom_full} for 2 QL Bi$_2$Te$_3$ and 5 QL Sb$_2$Te$_3$.

\section{Comparison with the bulk Hamiltonian}
\label{app:3D}

We also evaluate the FOM using the bulk Hamiltonian of Eq.~\eqref{eq:H_3D} for the materials Bi$_2$Se$_3$ and Bi$_2$Te$_3$, using the model parameters derived in Refs.~\cite{Liu2010B} and \cite{Zsurka2024}, respectively, with the results shown in Fig.~\ref{fig:fom_full_3D}. In the case of Bi$_2$Se$_3$ we observe a qualitatively similar behavior for FOM$_\text{a}$, with a peak appearing around $M_z = 22\,\text{meV}$ for 3 QL, probably matching the surface-state hybridization energy. For larger thicknesses ($4-6$ QL), the peak appears at lower magnetization, at the same $M_z$ for all three cases, approximately aligning with the confinement energy. Remarkably, below the DP we do not see a suppressed FOM$_\text{b}$ and the curves are almost identical to the ones obtained for FOM$_\text{a}$. For Bi$_2$Te$_3$ we observe again similar results above and below the DP. Surprisingly, when evaluated with the 3D model, the FOM in the $m_0>0$ case (5 QL) is not substantially worse than in the $m_0<0$ cases (3, 4, 6 QL). Similarly to the FOM determined with the thin-film Hamiltonian, in the case of 2 QL, there's no TR and thus no FOM, as the edge states are buried in higher-energy surface states. At high values of $M_z$, in all calculations the system enters the QAHI phase, with an exponentially suppressed PG and consequently a low FOM.

The difference in FOM obtained with the two Hamiltonians can be attributed to the fact that we have chosen the material parameters of the bulk Hamiltonian [see Eq.~\eqref{eq:H_3D}] on the basis of the size and topology of the hybridization gap in the thin-film limit~\cite{Zsurka2024}. These parameters do not guarantee an accurate description of the low-energy electronic structure (i.e. the Dirac cone). Consequently, when it comes to the bulk Hamiltonian with the material parameters chosen here, one should not expect to accurately capture such effects as the electron-hole asymmetry of the Dirac cone, which as we have seen previously, is critical in determining the FOM. Nonetheless, there is rather good overall agreement.

\section{List of abbreviations}
\label{app:abb}
For convenience, we provide a list of abbreviations used in the text in Table~\ref{tab:abbreviations}.

\begin{table}[t]
\begin{tabular}{ c | l } 
 \hline
 \hline
 TI & (three-dimensional) topological insulator \Tstrut\\
 MTI & magnetic topological insulator \\ 
 NR & magnetic topological insulator nanoribbon \\
 PNR & magnetic topological insulator nanoribbon \\
  & with proximity-induced superconductivity \\
 MBS & Majorana bound state \\
 DP & Dirac point \\
 QL & quintuple layer \\
 QAH(I) & quantum anomalous Hall (insulator) \\
 NI & normal insulator \\
 QSH(I) & quantum spin Hall (insulator) \\
 BdG & Bogoliubov-de Gennes \\
 PG & proximity-induced superconducting gap \\
 TR & (single-subband) topological regime \\
 LES & low-energy state \\
 SSG & surface-state gap \\
 FOM & figure of merit \\
 \hline
 \hline
\end{tabular}
\caption{List of abbreviations.}
\label{tab:abbreviations} 
\end{table}

\bibliographystyle{apsrev4-2}
\bibliography{main.bib}

\end{document}